\documentclass[10pt,letterpaper]{article}
\usepackage[top=0.85in,left=0.75in,footskip=0.75in]{geometry}
\usepackage{changepage}
\usepackage[utf8]{inputenc}
\usepackage{textcomp,marvosym}
\usepackage{fixltx2e}
\usepackage{amsmath,amssymb}
\usepackage{cite}
\usepackage[right]{lineno}
\usepackage{microtype}
\DisableLigatures[f]{encoding = *, family = * }
\usepackage{rotating}
\usepackage{float}

\usepackage[aboveskip=1pt,labelfont=bf,labelsep=period,justification=raggedright,singlelinecheck=off]{caption}
\makeatletter
\renewcommand{\@biblabel}[1]{\quad#1.}
\makeatother

\date{}

\usepackage{rotating}
\usepackage{graphicx}
\usepackage{amssymb,amsfonts,amsmath}
\usepackage{bm}
\usepackage[normalem]{ulem}
\usepackage{morefloats}
\usepackage{bbm}
\usepackage{dsfont}
\usepackage{color}

\def\beq{\begin{equation}}
\def\eeq{\end{equation}}
\def\bea{\begin{eqnarray}}
\def\eea{\end{eqnarray}}

\newcommand{\phat}{\hat{\mathbf p}}
\newcommand{\vel}{{\mathbf v}}

\begin{document}
\vspace*{0.35in}

\begin{flushleft}
{\Large
\textbf\newline{Coherent motion of monolayer sheets under confinement and its pathological implications}
}
\newline
\\
S S Soumya\textsuperscript{1},
Animesh Gupta\textsuperscript{2},
Andrea Cugno\textsuperscript{3},
Luca Deseri\textsuperscript{3},
Kaushik Dayal\textsuperscript{4},
Dibyendu Das\textsuperscript{2},
Shamik Sen\textsuperscript{5},
Mandar M. Inamdar\textsuperscript{1,*}
\\
\bigskip
\bf{1} {Department of Civil Engineering,
Indian Institute of Technology Bombay, Mumbai 400076, India}.
\\
\bf{2} {Department of Physics,
Indian Institute of Technology Bombay, Mumbai 400076, India}.
\\
\bf{3} {DICAM-Dept. of Civil, Environmental 
and Mechanical Engineering, University of Trento, via Mesiano 77, 38123 Trento-Italy}.
\\
\bf{4} {Civil and Environmental Engineering
Carnegie Mellon University
Pittsburgh PA 15213}.
\\
\bf{5} {Department of Biosciences and Bioengineering,
Indian Institute of Technology Bombay, Mumbai 400076, India}.
\\
\bigskip

* minamdar@civil.iitb.ac.in

\end{flushleft}
\section*{Abstract}

Coherent angular rotation of epithelial cells is thought to contribute to many vital physiological processes including tissue morphogenesis and glandular formation. However, factors regulating this motion, and the implications of this motion if perturbed remain incompletely understood. In the current study, we address these questions using a cell-center based model in which cells are polarized, motile, and interact with the neighboring cells via harmonic forces. We demonstrate that, a simple evolution rule in which the polarization of any cell tends to orient with its velocity vector can induce coherent motion in geometrically confined environments. In addition to recapitulating coherent rotational motion observed in experiments, our results also show the presence of radial movements and tissue behavior that can vary between solid-like and fluid-like.  We show that the pattern of coherent motion is dictated by the combination of different physical parameters including number density, cell motility, system size, bulk cell stiffness and stiffness of cell-cell adhesions. We further observe that, perturbations in the form of cell division can induce a reversal in the direction of motion when cell division occurs synchronously. Moreover, when the confinement is removed, we see that the existing coherent motion leads to cell scattering, with bulk cell stiffness and stiffness of cell-cell contacts dictating the invasion pattern. In summary, our study provides an in-depth understanding of the origin of coherent rotation in confined tissues, and extracts useful insights into the influence of various physical parameters on the pattern of such movements.

\section*{Introduction}
Collective cell migration is central to both physiological processes such as morphogenesis and wound healing, and pathological processes like cancer invasion~\cite{martin2004parallels, block2004wounding, rorth2009collective, rorth2012fellow, friedl2012classifying, kaunas2014cell}. Epithelial and endothelial cells collectively migrate in intricate patterns within a tissue by virtue of their adhesion to their neighboring cells and to the extracellular matrix (ECM)~\cite{ilina2009mechanisms, ng2012substrate}. Further, on 2D confined geometries, these cells exhibit coherent angular movement (CAM)~\cite{doxzen2013guidance, huang2005symmetry, brangwynne2000symmetry, tchao1982}. Interestingly, such coordinated movements have also been documented in various \textit{in vivo} processes including egg chamber elongation in {\it Drosophila}, ommatidial rotation in {\it Drosophila} and development of spherical mammary acini~\cite{haigo2011global,cetera2015round, cetera2014epithelial,tanner2012, wang2013,mirkovic2006cooperative}. In addition to these types of tissues, such large scale rotations are also observed in confined dictyostelium colonies and bacterial suspensions~\cite{rappel1999, goldstein2013}.  Moreover, non-living, active materials such as vibrated, granular materials also exhibit spontaneous CAM when confined~\cite{kumar2014}. Thus, large scale rotational movements under confinement are ubiquitous in `active systems' -- both non-living and living.\\

Active systems have been modeled using a variety of approaches ranging from discrete, self-propelled particle modeling (SPM) to active hydrodynamical theories~\cite{vicsek2012, marchetti2013}.  Of special interest are theories, which involve discrete or continuum elements with self-propulsion, and are successfully used to describe collective motion in epithelia~\cite{deforet2014, doxzen2013guidance, li2014coherent, szabo2006phase, Lee2011, szabo2010, basan2013alignment}. The common thread connecting these diverse modeling attempts is the presence, in some form, of self propulsion velocity $v_0$ originating from actin polymerization and polarization $\hat{{\mathbf p}}$ for the active elements, in addition to the elastic and viscous interactions of the elements with their surrounding constituents. The polarization $\hat{{\mathbf p}}$ is a coarse-grained representation of front-rear asymmetry of a migrating cell resulting from various factors, e.g., Rho GTPase gradient~\cite{levine2014} and position of centrosome in relation to the nucleus\cite{desai2009cell,huang2014polarized}. A SPM-based cellular Potts model has successfully replicated the existence of CAM in confined epithelia~\cite{doxzen2013guidance}. Similarly, a recent study has also demonstrated that a particle based model for confined epithelia, where cells are represented as self-propelled points connected to their neighbors with elastic springs, also gives rise to CAM~\cite{li2014coherent}. A distinct, but related, formalism that utilizes dissipative particle dynamics for sub-cellular components, has been used to demonstrate spontaneous rotation of two tightly connected, and confined cells~\cite{leong2013}. Recently, Camley et al. used a phase-field method for studying the emergence of coherent rotation in a pair of cells confined to adhesive micropatterns, and demonstrated that even subtle differences in cell polarity mechanisms (via contact inhibition, neighbor alignment, velocity alignment, etc) greatly influence the pattern of collective movement~\cite{levine2014}.

Though SPM has also been utilized to represent stable vortex formation in confined bacterial suspensions and driven granular media~\cite{goldstein2013, kumar2014}, the mechanism of vortex formation relies on hydrodynamic coupling between the active particles through the surrounding media. This is distinct from the collective behavior in epithelia, in which the tight inter-cellular contacts play a crucial role in tissue movements~\cite{rorth2012fellow, doxzen2013guidance}.\\

Despite the presence of several models addressing CAM in epithelia, a number of crucial questions still remain unanswered. For example, there is no simple understanding as to why CAM spontaneously emerges in such systems. Additionally, the diversity of experimental findings from similar experiments with near-identical setup~\cite{doxzen2013guidance, deforet2014, radler2015} raises the possibility that small perturbations in physical parameters associated with the experiments are likely to influence the various hydrodynamic modes exhibited by cells, and hence motivates the necessity of identifying some of the critical physical parameters. Though the confinement provided for {\it in vitro} cultures arises quite naturally on the micropatterned geometries, the confinement for epithelia  {\it in vivo} comes from being embedded in a larger tissue~\cite{kabla2014, kabla2012collective}. What role the nature of confinement plays on the emergence and sustenance of CAM is another issue that needs addressing. Similarly, the influence of confinement geometry on CAM is also not clear, and is particularly relevant to several \textit{in vivo} situations. For example, the annular geometry is the simplest non-convex geometry that is relevant in understanding CAM in biological lumens \cite{friedl2009collective}. Though CAM on this geometry has been addressed before~\cite{li2014coherent}, the role of cell cohesivity on the nature of CAM still remains unaddressed. Finally, the roles of internal and external perturbations mimicking various \textit{in vivo} processes, in the form of cell division and the loss of confinement on CAM also remain unknown. In this  paper, we employ SPM using cell center representation of cells to computationally answer these questions~\cite{szabo2006phase, li2014coherent}. Additionally, using simple calculations we also provide analytical insights to get a better understanding of CAM in epithelia. We show that the nature of coherent motion can be both solid-like or fluid-like, and is dictated by the combination of different physical parameters including number density, cell motility, system size, bulk cell stiffness and stiffness of cell-cell adhesions. Our results also predict that synchronous cell division can lead to change in the direction of rotation for CAM in annular geometries. Finally, we show that depending on the properties of cells and cell-cell adhesions, CAM leads to different patterns of invasion when the confinement is removed. Collectively, our results illustrate the influence of cell density, system size, cell motility, cell division, and stiffness of cell and cell-cell adhesions in regulating CAM.

\section*{Computational Model}

An epithelial sheet is comprised of a group of cells that are connected to each other via cadherin bonds to form a monolayer. Many experimental observations have demonstrated that cells in this network are persistently motile, and upon reaching a critical density show collective migration behavior\cite{poujade2007collective}. Presence of front-rear polarity axis is known to be essential for migrating cells. This polarity axis manifests in migrating cells in different forms like: (i) increased actin activity in the front and formation of actin structures such as lamellipodia, (ii) localization of the microtubule organizing center (MTOC) at the front of the nucleus with microtubule growth towards the leading edge, (iii) gradients in cell-ECM adhesion, and (iv) establishment of front-rear gradients in the activity of GTPases such as Rac/Cdc42~\cite{frontrear2015}. Cell polarity is actively maintained and constantly steered by complex mechano-chemical processes governed by cell-cell and cell-ECM interactions~\cite{collins2015, cai2014}. A surprisingly simple upshot of these complex processes in terms of mechanical observables is that, in epithelial sheets such as MDCK tissue,  the polarization of constituent cells is closely oriented with the principal direction of stress as well as with their average velocity~\cite{das2015, tambe2011collective}. Keeping these experimental observations in mind, we have utilized a simple model to explore how mechano-chemical properties of individual cells impact their collective behavior in confined epithelial sheets.

\begin{figure}[H]
\centering
\includegraphics[width=11 cm]{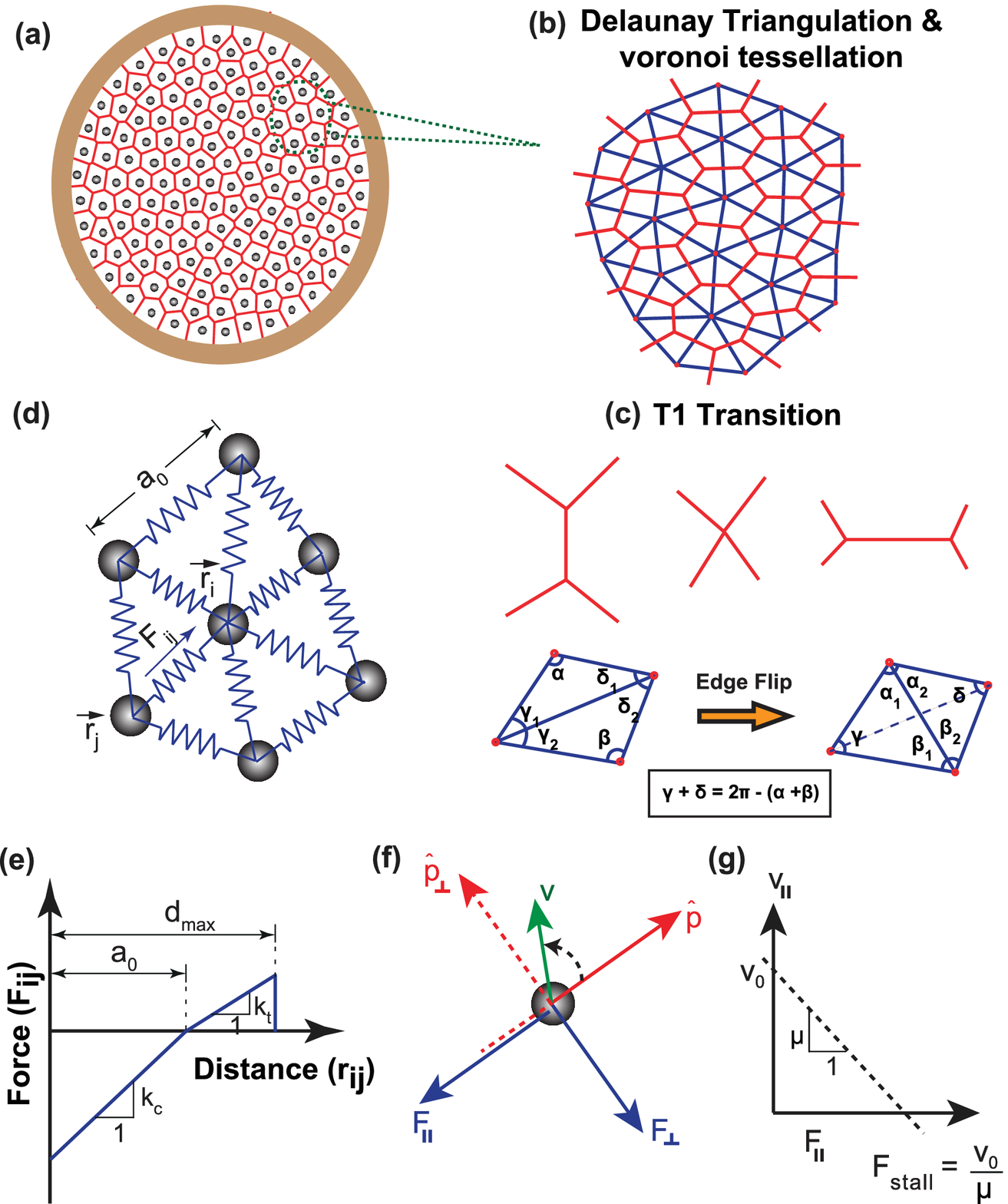}
\caption{\label{fig:1} {\bf A schematic of cell center model depicting the arrangement of cells and the forces acting on them}. (a) A 2-D monolayer of epithelial cells, confined inside a circular geometry is considered with cells represented as points at their center. (b) Delaunay triangulation (blue) has been used to model cell - cell connectivity, which finds the nearest neighbors of each point and form the connectivity array accordingly. Because of the greater clarity it affords and better connection with the experimental geometry, Voronoi tessellation (topological dual of Delaunay triangulation) is used for visualization of cells. (c) When two originally connected cells move apart and form new neighbors, the connectivity of the system is updated using Delaunay triangulation. This connectivity update automatically takes T1 transitions into account. (d) Enlarged view of a representative cell $i$, along with its connection to neighboring cells. The position vector of this cell center is denoted by $r_i$ and position vector of its $j^{th}$ neighbor is denoted by  $r_j$. The blue arrow indicates the force, $F_{ij}$ acting between cells $i$ and $j$. The total force acting on $i^{th}$ cell is the sum of the contributions from all the connecting neighbors. (e) The interaction between two adjacent cells is either compressive or tensile, depending upon the relative deformation of connecting spring with respect to its undeformed length, $a_0$. Here compressive and tensile stiffness of each spring is represented by $k_c$ and $k_t$, respectively. While $k_c$ mimics the bulk cell stiffness, $k_t$ mimics cell-cell cohesivity. It is assumed that if the deformation of any spring is greater than $ d_{\rm max}$, the cell-cell connection is broken and there is no force transfer between these two cells. (f) Force acting on each cell is resolved along anti-parallel ($F_{\rm ll}$) and perpendicular($F_\perp$) to the direction of the cell's polarization($\hat {\bf p}$). Here $v$ denotes the velocity vector on each particle. (g) Velocity profile in the direction of polarization as a function of $F_{\rm ll}$.}
\end{figure}

For modeling the  collective mechanics of cells, we have adopted a `cell center-based mechanics model' with cells represented as discrete points at their center of mass~\cite{szabo2006phase,pathmanathan2009computational,marmaras2010mathematical}. As shown in Fig.~\ref{fig:1}, the whole epithelial tissue is represented as a continuous sheet with cell-cell cadherin junctions represented by simple harmonic springs~\cite{bameta2012broad, szabo2006phase}. Each cell is assumed to exert an attractive or repulsive force on its neighboring cells depending on the relative deformation of springs with respect to their undeformed length, $a_0$ and stiffness, $k$. The force acting on any cell at any time, $t$, is the sum of the contributions of all the connecting neighbors. Thus, if $\bm r_i$ represents the position of $i^{\rm th}$ cell, the net force exerted on that cell by neighbors ($m$, say) is given by
\begin{equation}
\bm F_i = \sum \limits_{j \in{\rm neighbor}}k(|{\bm r_j - \bm r_i}|-a_0)\bm e_{ij}
\label{eq:force}
\end{equation}
where, $\bm e_{ij}= \frac{(\bm r_j- \bm r_i)}{|\bm r_j-\bm r_i|}$ represents the unit vector along the direction connecting the $i^{\rm th}$ cell with its $j^{\rm th}$ neighbor. Depending on the relative deformation of springs with respect to the natural length, the interaction potential can either be tensile or compressive. In order to avoid force transfer between distant neighbors, it is assumed that when the deformation of spring is greater than a threshold, $d_{\rm max}$, no force transfer occurs between those two cells. For all our simulations, we took the value of $d_{\rm max}$ equal to $1.3~ a_0$~\cite{szabo2006phase}. Thus the value of spring stiffness for the entire range of deformation can be written as:
\begin{equation}
  k=\begin{cases}
    0, & \text{if $(|{\bm r_j - \bm r_i}|-a_0 )> d_{\rm max}$}.\\
    k_t, & \text{if $0\leq (|{\bm r_j - \bm r_i}|-a_0) \leq d_{\rm max}$}.\\
    k_c, & \text{if $(|{\bm r_j - \bm r_i}|-a_0)\leq 0 $}.
  \end{cases}
\end{equation}
In the above expression, $k_c$ and $k_t$ represent the bulk cell stiffness and the stiffness of cell-cell adhesions (or cohesivity), respectively. Figure~\ref{fig:1}(e) illustrates the attractive/repulsive force acting on each cell. The cells are allowed to exchange their neighbors, which are obtained by repeated Delaunay triangulation~\cite{pathmanathan2009computational, li2014coherent}. For a given set of cell centers, Delaunay triangulation provides a connectivity for cells that produces the least number of distorted triangles, i.e., triangles with least shear strain. Delaunay triangulations are dual to Voronoi tessellations (Fig.~\ref{fig:1} (b), (c)) and the Voronoi polygon for a given cell center can be modeled to be the cell itself (see Materials and Methods).

In our model, cells are assumed to act as self propelled active particles~\cite{szabo2006phase}, with their inherent motility ($v_0$) representing the speed with which they move in the absence of any external force. The preferential direction of cell's motion (i.e., polarization) is represented by the vector $\hat{\bm p_i}$, which is a coarse-grained representation of the front-rear polarization in a motile cell~\cite{frontrear2015}.
As cells move over a viscous substrate with mobility $\mu$, the drag force acting in the opposite direction of motion balances the internal forces. If $\bm r_i$ is the position vector of $i^{\rm th}$ cell, its velocity at time $t$ can be written as:
\begin{equation}
\bm v_i=\frac{d\bm r_i}{dt}=v_0\hat{\bm p_i}+\mu \bm F_i
\label{eq:velocity}
\end{equation}
Similar to the procedure followed elsewhere~\cite{szabo2006phase} and as motivated earlier, we assume that the cell's polarization vector tends to orient with its velocity vector as per the following equation:
\begin{equation}
\frac {d \hat {\bm p}_i}{dt}=\xi(\hat {\bm p}_i \times \hat {\bm v}_i . \bm {\hat e_z})\hat {\bm p}_{i}^\perp
\label{eq:polarisation}
\end{equation}
where $\hat {\bm v}_i$ is the unit velocity vector and $\bm {\hat e_z}$ is the unit vector perpendicular to the plane. The parameter $\xi$ represents the polarization coordination constant determining the tendency of cell's polarization to rotate and align with the velocity vector. We do not account for noise in our simulations~\cite{romanczuk2012} as we are primarily interested in a mean-field understanding of CAM, and noise is known to typically increase fluctuations in the system~\cite{li2014coherent}.

\subsection*{Numerical estimates of parameters used in the study}
Before employing our model for any qualitative predictions, it is essential to estimate the real values of various parameters used in the study. Consistent with previous studies~\cite{ Lee2011, bameta2012broad,petitjean2010velocity}, $20~\mu{\rm m}$ and $20~\mu{\rm m/hr}$ were taken as cell length and cell speed, respectively. Unless specified, for all the simulations, the non-dimensional value of $\mathrm {a_0}$ and $\mathrm{v_0}$ were taken as $1$. Assuming a substrate drag coefficient ($\mathrm{\zeta}$) of $\mathrm{100\ pN~hr/\mu m^3}$~\cite{Lee2011}, the value of mobility $\mathrm{\mu}$ was calculated to be $\mathrm {1/(\zeta a_0^2)=2.5 \times 10^{-5}}$ $\mathrm{\mu m/pN~hr}$. Length, time and force are expressed in units of $a_0 = 20$~$\mu$m, $\tau_0 = a_0/v_0 = 1$ hr and $f_0 = v_0/\mu = 8 \times 10^5$~pN, respectively. The non-dimensional value of mobility ($\mathrm{\bar \mu}$) was taken as 1 for all the simulations unless otherwise specified. The value of stiffness of cell-cell connection is given by the expression $k = E~h/ (2\sqrt 3~(1-\nu))$~\cite{boal2012mechanics}  where $h$, $E$ and $\nu$ represent the height, Young's modulus and Poisson's ratio of the cell, respectively  (see supplementary information (SI) Text~S1 for derivation). Assuming values of $E=10~$kPa, $h = 5~\mu$m and $\nu = 0.5$~\cite{tambe2011collective}, the value of stiffness was estimated to be $k \approx 0.03$~N/m. In our simulations, we have used the non-dimensional value of stiffness ($\mathrm{\bar k}$) in the range of $1-10$. For this range, the real value of stiffness was calculated as $k = \bar{k} v_0/\mu a_0$ yielding a value of $k= 0.04 - 0.4$~N/m, which is close to the actual value of cell-cell stiffness derived above. Due to uncertainty in the value of $\zeta$, the non-dimensional value of $k$ can indeed have a some variability.  The outer radius of substrate was taken as 100 $\mu$m for all the simulations~\cite{doxzen2013guidance}, unless stated otherwise. For annular geometry,  the inner radius was taken as 70 $\mu$m. The number of cells were varied between $100 -170$ for various simulations, yielding an average cell density in the range of 3000 - 6000~cells/${\rm mm}^2$ which closely matches with previous experimental studies~\cite{doxzen2013guidance}.
\section*{Results}
\subsection*{Coherent rotation of cells confined in circular geometry}
\begin{figure}[H]
\centering
\includegraphics[width=11.4 cm]{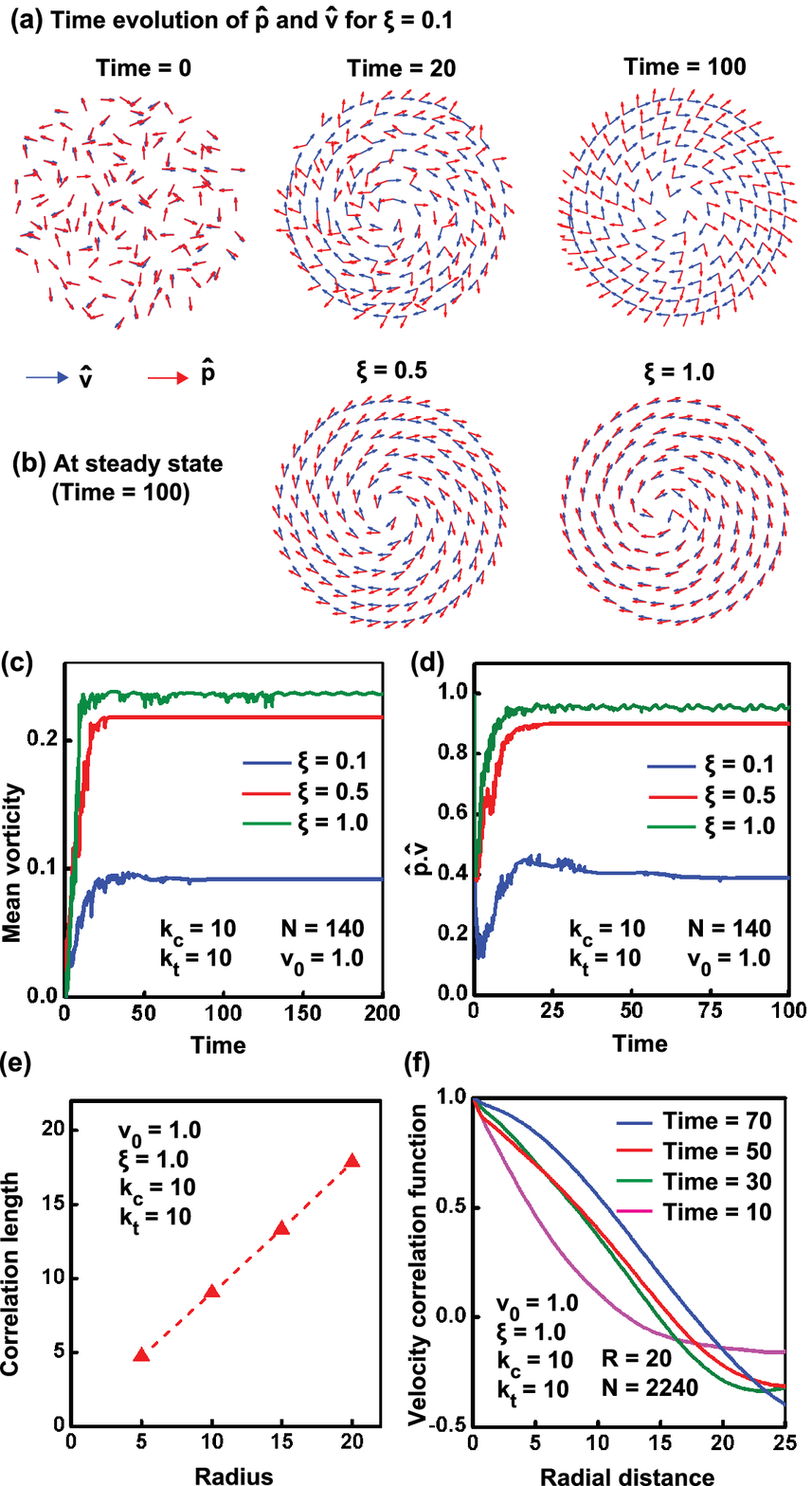}
\caption{\label{fig:2} {\bf Coherent rotation of cells on circular geometry.} {(a) The time evolution of polarization vector, $\hat {\bf p}$ and velocity vector $\hat {\bf v}$ is shown for $\xi = 0.1$. The evolution rule for polarization is chosen in such a way that, from an initial random orientation, $\hat {\bf p}$ will try to orient along velocity vector with time. (b) The coordination between $\hat {\bf p}$ and $\hat {\bf v}$ is decided by the parameter $\xi$.  The higher the value of $\xi$, higher is the tendency of $\hat {\bf p}$ to orient along $\hat {\bf v}$. The orientation of $\hat {\bf p}$ and $\hat {\bf v}$ at steady state for $\xi=0.5$ and $\xi=1$ are also shown. (c) Mean vorticity for systems with different $\xi$ is plotted as a function of time. (d) The tendency of polarization vector to orient with velocity vector is shown by the plot between $\hat {\bf p}.\hat {\bf v}$ and time. As the value of $\xi$ increases, value of $\hat {\bf p}.\hat{\bf v}$ approaches 1, indicating perfect alignment between two vectors. (e) A plot of velocity correlation length for varying system size shows that correlation length equal to the confinement size. (f) A plot of correlation function with time shows that the velocity correlation length increases with time, till the coherent rotation sets in.}
}
\end{figure}

Various theoretical studies modeling the behavior of cells on micro-patterned substrates have established the emergence of coherent rotation of cells under confined conditions~\cite{li2014coherent, szabo2006phase}. Similar to these studies, our model also shows the emergence of a persistent mode of rotation for a group of cells ($N = 140$) when confined on a circular substrate ($k_c=k_t=10$, $\xi =1$, $v_0=1$, $\mu = 1$) (Video S1). While the theory of active elastic systems attributes the onset of rotational motion to energy transfer to the lowest modes~\cite{ferrante2013collective, ferrante2013PRL}, a systematic analysis of this phenomenon in the context of epithelial sheets remains to be performed. Using our model, we demonstrate that rotation is indeed the preferred mode of motion for tissues confined in circular geometries---this mode of CAM is very different than that observed in bacterial suspensions~\cite{goldstein2013} (also see SI Text S2). Figure~\ref{fig:2}(c) illustrates the quantification of this rotational motion in terms of mean vorticity of the system (See materials and methods). After an initial transient mode, cells start to rotate steadily as evidenced by the constant value of the mean vorticity of the system. The onset of rotation depends on the parameter $\xi$, which reflects the tendency of the cell's polarization to orient along its velocity (Fig.~\ref{fig:2}(a)). The greater the value of $\xi$, higher is the tendency of polarization vector to reorganise and align along the velocity vector, resulting in faster initiation of coherent rotation of cells (Fig.~\ref{fig:2}(b)). Figure~\ref{fig:2}(d) emphasizes this by plotting the scalar product of polarization vector and velocity vector ($\hat {\bf p}.\hat {\bf v}$) as a function of time. From the figure it is seen that, as the value of $\xi$ increases, coordination between $\hat{\bf p}$ and  $\hat{\bf v}$ is builds up faster resulting in a faster approach to steady state of motion. We would also like to emphasize that, for larger values of $\xi$, the time scale for polarization evolution can be faster than the relaxation of a few long wavelength radial modes (see SI Section Text~S2 and SI Video~S3). In this case,  some long wavelength radial modes can be sustained during the coherent rotation and the tissue can exhibit radial movements that are similar to those observed by  Deforet et al.~\cite{deforet2014}.  Additionally,  as the confinement radius $R$ for the tissue increases, these radial movements become prominent even at lower values of $\xi$ ( Section Text~S2 of SI and SI Video S19). This is because, larger the system size, lower is the stiffness of long wavelength radial modes, and hence slower is their decay. This behavior of increasing radial velocity for the tissue with increasing confinement size is also observed by Deforet et al. in their experiments (see SI Fig.~4 of Ref.~\cite{deforet2014}).

It was reported by Doxzen et al that, for tissues with confinement size greater than the velocity correlation length ($\approx 200~\mu$m), there was no onset of CAM within the observation window of around $48$ hours~\cite{doxzen2013guidance}. However, we find from our simulations that irrespective of tissue size ($R$), the tissue always reaches the steady state of coherent rotation (see Fig.~\ref{fig:2}). In other words, we find that the steady state velocity correlation length is set by the size of the confined tissue. However, the time required to reach the steady state is higher for larger tissues (see Figs.~\ref{fig:2}(e)--(f) and SI Fig.~S5). This increase in the time required to reach the steady state may be attributed to the presence of a greater number of long wavelength modes for the larger system, as described above. The presence of these modes would interfere with the transfer of cellular motility to the rotational mode.
We can reconcile our simulation results with the experimental observations by noting that, as the time required for setting the coherent motion is greater for larger tissues, the tissue is likely to be perturbed by certain unknown factors (e.g., cell proliferation) in that additional time. The resulting mechanical and polarization perturbations may, therefore, further delay the onset of coherence with respect to the experimental time window, or make CAM infeasible. We predict that in the absence of perturbations, even a large confined tissue can undergo CAM. These predictions differ from the observation of finite velocity correlation lengths of around $10$ cell lengths in unconfined tissues (e.g. Refs.~\cite{petitjean2010velocity, das2015}), wherein different boundary conditions (e.g., leader cells, high cable tension, etc) are likely to lead to qualitatively different behavior from that of confined tissues.

Collectively, these results illustrate the effect of confinement in inducing coherent angular motion. Under \textit{in vivo} conditions, such confinement may be provided by non-motile cells~\cite{kabla2014} possessing higher substrate frictions than motile cells (see SI Text~S4 and SI Figs. S3, S4, Video S4--S11). Under these conditions, the efficiency of coherent motion is dictated by the ratio of substrate frictions between the two cell types.

\subsection*{Cell crowding leads to fluidisation of tissue}
\begin{figure}[H]
\centering
\includegraphics[width=11.4 cm]{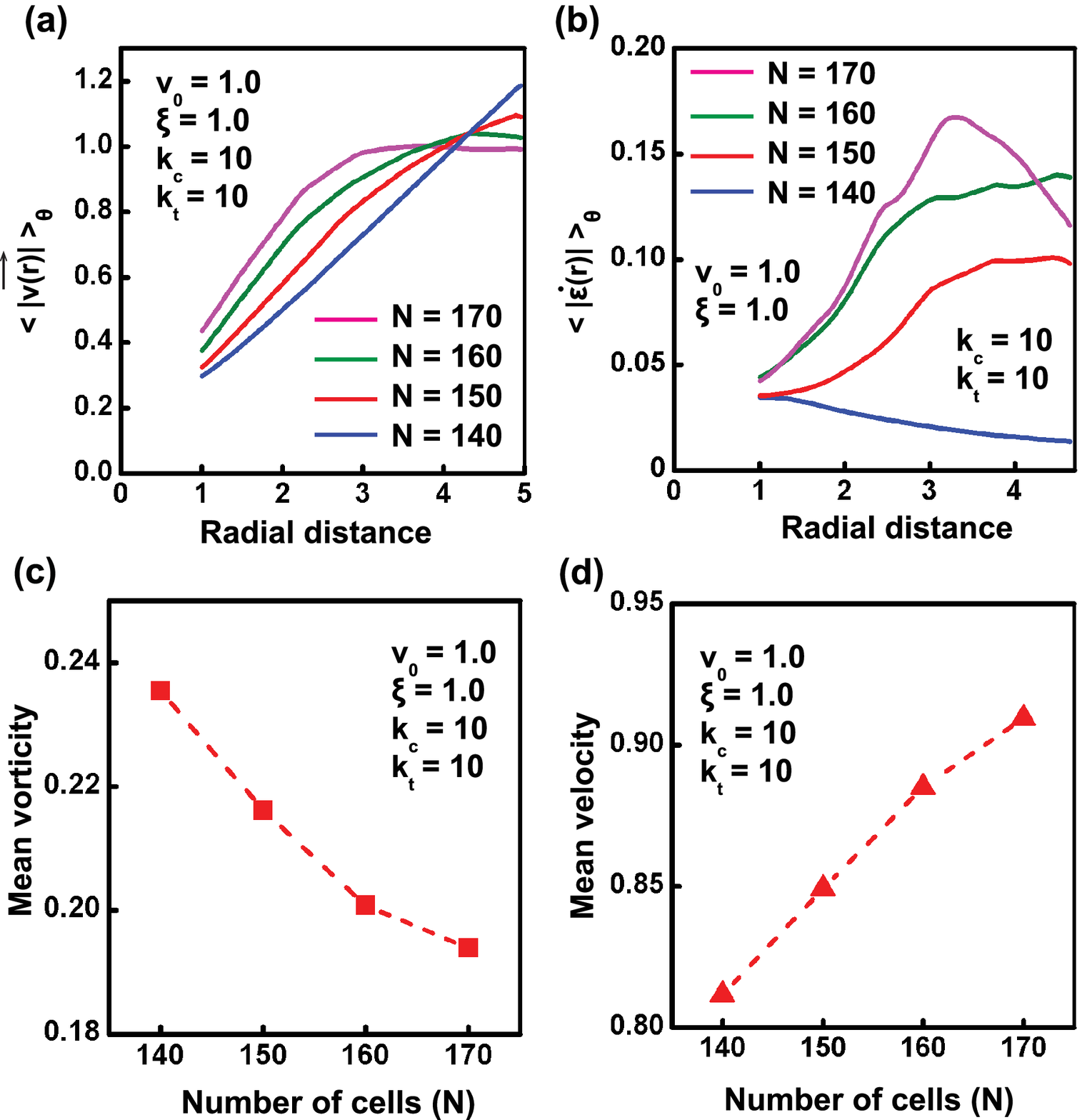}
\caption{\label{fig:3} {\bf Cell crowding leads to fluidisation of tissue.} {(a) The relationship between velocity and radial distance is examined for varying number density. Keeping the values of other parameters same as in previous simulations, the absolute velocity, $|v|$ averaged over time, after the system reaches steady state,} is plotted as a function of radial distance for varying number of cells $N= \{140,150,160,170\}$. As the number density of system increases, the velocity-radial distance curve become less linear, indicating the presence of shear in the system. (b) Variation of principal shear strain rate along the radial distance plotted as a function of number density. Increase in shear rate with number density illustrates the fluidisation of tissue induced by cell density. (c) Vorticity of system decreases with increase in cell density. (d) Without considering the effect of contact inhibition, mean velocity of the system increases with number density.}

\end{figure}

As the presence of a rotational mode of migration under confinement is well established by now, we focused our attention in understanding the characteristics of that motion in detail. Studies by Doxzen {\it et. al.} have shown that the movement of small circular tissues under confinement is similar to solid body rotations with angular velocity $\omega$ equal to $\frac{4v_0}{3R}$, where $R$ is the radius of circle~\cite{doxzen2013guidance}. Further, the linear relationship between velocity and radial distance for rotating cell collectives obtained by multiple research groups support the argument of solid body rotations~\cite{doxzen2013guidance, li2014coherent}. However, what factors influence this solid-like tissue behavior has not been addressed. Here, we show that cell density is one such parameter dictating the nature of tissue behavior. As shown in Video S1, at lower cell densities, system behaves as an elastic solid with negligible neighbor changes and a linear velocity versus radial distance relationship (Fig.~\ref{fig:3}(a)). Increase in number of cells in the system while keeping the size $R$ constant, i.e., increase in cell density, leads to an interesting phenomena. Increase in cell density alters the nature of the velocity versus radial distance relationship and induces a transition from solid-like behavior ($N = 140$) to that like a fluid ($N = 170$). Specifically, with increase in cell density, the linear velocity versus radial distance curve becomes more saturating. At the highest cell density ($N=170$), the velocity plateaued to $v_0 = 1$ at the edges. One of the probable reasons for this change is the large shear that the system experiences at such densities, as evident from the relative sliding of cells past each other (Video S2). Quantification of the shear strain rate ($\dot \epsilon_{xy}$) from the rate of deformation tensor as $\dot \epsilon_{xy}=\frac{1}{2}\left(\frac{\partial u}{\partial y}+\frac{\partial v}{\partial x}\right)$ was performed to obtain additional insight into the magnitudes of shear experienced by the cells at various cell densities. A plot showing the variation of principal shear strain rate as a function of radial distance shows that with increase in cell number, the shear in the system also increases (Fig.~\ref{fig:3}(b)). Collectively, the above numerical results indicate that the number density of cells alters the behavior of system; i.e, at lower cell densities, system behaves like an elastic solid and at higher cell densities, system becomes more fluid-like.

\begin{figure}[H]
\centering
{\includegraphics[width=11.4 cm]{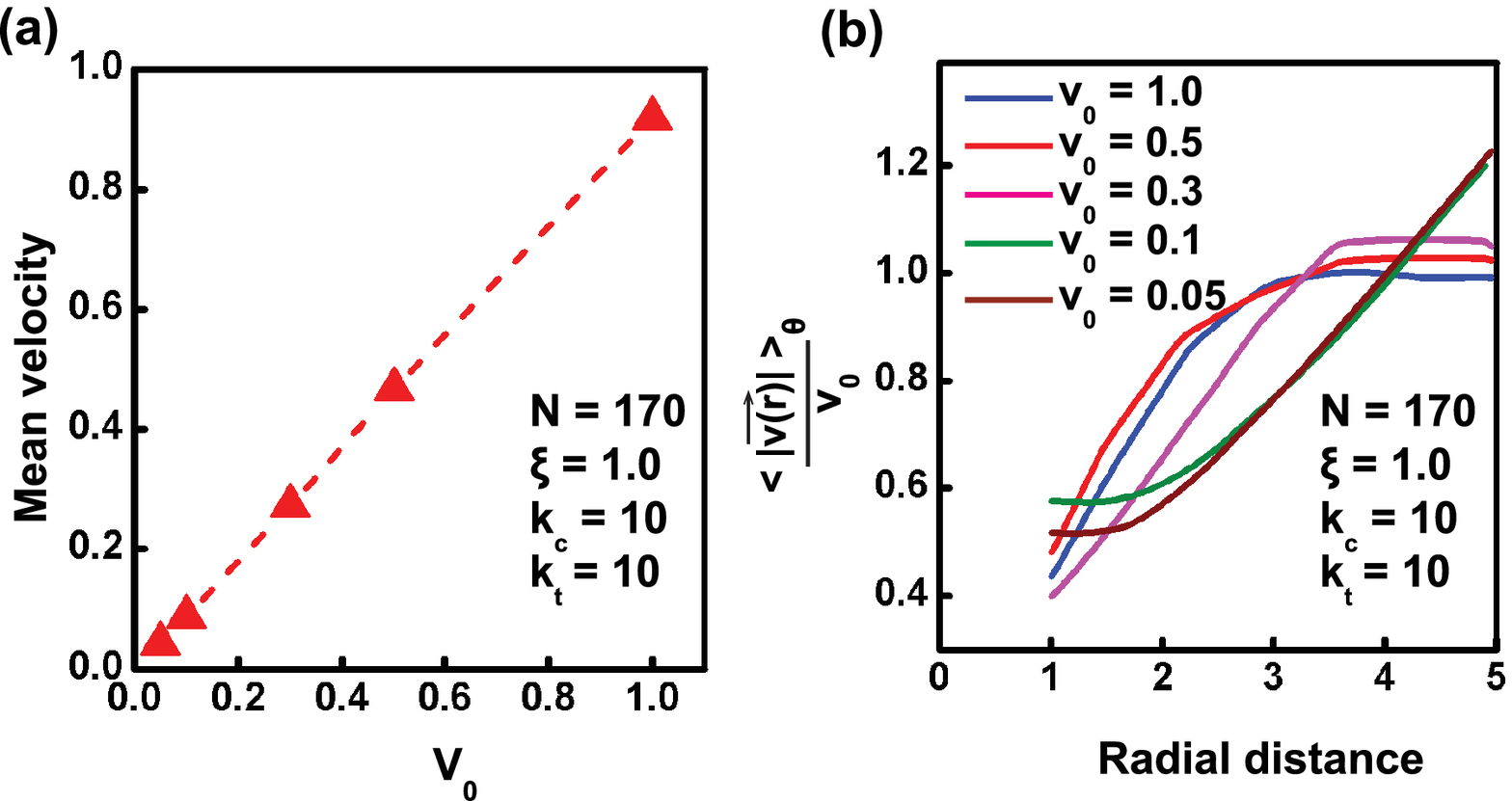}}
\caption{
\label{fig:4} {\bf Cell motility dictates the fluidized behavior of tissue.}} {(a) Mean velocity for varying values of cell motility ($v_0$). (b) Normalized velocity-radial distance plot for varying values of $v_0$ for $N = 170$.
}
\end{figure}

While studying the effect of cell crowding on the nature of coherent rotation, we assumed that the motile cell speed or the fraction of motile cells is not modified by cell density. Consequently, we find that the mean speed of the cells in the tissue increases with cell density (Fig.~\ref{fig:3}(d)). This finding follows from our observation in Fig.~\ref{fig:3}(a) wherein upon increase in cell density, the tissue fluidises, as a result of which more and more layers of the tissue move with speeds comparable to $v_0 = 1$. On the other hand, when the tissue behaves elastically (for $N = 140$), the tissue rotates as a rigid body with cell speed comparable to $v_0$ at the edges, but significantly lower speed of cells in the interior. However, while studying the effect of cell density on velocity profile of the over-confluent tissue, the condition of contact inhibition observed experimentally~\cite{mayor2010} has not been taken into account. To mimic the condition of contact inhibition for a denser system, and reconcile the experimental observations of decrease in mean velocity with increase in number density~\cite{doxzen2013guidance}, we have considered the following cases: (i) due to crowding, the self-propelled speed of cells can be smaller on account of cells forming smaller lamellipodia~\cite{doxzen2013guidance}  (see Fig.~\ref{fig:4}(a)), or (ii) due to crowding, a fraction of cells are possibly not motile (see SI Fig.~S6). Both of these effects are feasible due to contact inhibition of motility in crowded tissues. For both cases, as expected, we observed reduction in mean cell speeds. Additionally, we can also see from Fig.~\ref{fig:4}(b) that the tissue shows fluidisation for value of $v_0$ as low as $0.3$; only at really low $v_0 = 0.1$ does it recover back its elastic behavior. Thus, for appropriate values of $v_0$ at large $N$, we can observe lower mean cell speeds, concurrently with a fluid-like behavior for the overall tissue.

\subsection*{Exact steady state solution when the tissue is a linear, isotropic, homogeneous, elastic solid}

It appears from the above findings that $N$ and $v_0$ are two  parameters that tune the solid to fluid transition in coherently rotating tissues. In order to gain insight into this transition, we first analytically obtain mechanical steady state of the tissue by modeling it as a homogeneous, linear, elastic solid. This continuum description seems reasonable when cell-cell connectivity in the tissue is maintained during coherent rotation~\cite{boal2012mechanics}. For such a tissue  constrained within a circular patch and with zero tangential tractions, we can obtain \emph{one particular} steady state solution, based on the following assumptions:
\begin{enumerate}
\item The solution is radially symmetric, i.e., there is no $\theta$ dependence
\item The polarization $\phat$ is aligned in the $\theta$ direction
\item All cells are motile
\end{enumerate}
Due to ensuing radial symmetry, the confinement is expected to induce isotropic compression only, i.e., $\sigma_r = \sigma_\theta = p$, which should not interfere with  the shear strain (stress) in the tissue due to motile forces.

The easiest solution to visualize is a small elastic displacement $u_r$ and $u_\theta$ superimposed on a rigid body rotation with angular speed $\omega$. If this solution is indeed possible then, $\phat$ and cell velocities will be both aligned in the tangential direction. The continuum form of polarization evolution equation (see Eq.~S1) would be,
\bea
\label{eq:corot}
{D\phat \over Dt} = \xi (\phat \times {\mathbf v}).\mathbf{e}_z \phat_{\perp}
\eea
for the current model, where $D/Dt$ represents the co-rotational material derivative for the elastic sheet~\cite{chadwick2012, marchetti2013}. We look at the steady state solution when $\phat$ would not vary temporally.

The equation of equilibrium, respectively, in the radial and the tangential direction for the current elastic sheet with above conditions will be~\cite{sadd2009}:
 \bea
 \label{radial}
 {Eh \over 2 (1 + \nu)}\left ( {\partial^2 u_r \over \partial r^2} + {1 \over r} {\partial u_r \over \partial r} - {u_r \over r^2}\right ) + {Eh \over 2 (1-\nu)} {\partial \over \partial r}\left( {\partial u_r \over \partial r} + { u_r \over r}\right) &=& 0\\
 \label{theta}
 {Eh \over 2(1 + \nu)} \left( {\partial^2 u_{\theta} \over \partial r^2 } + {1\over r} {\partial u_{\theta} \over \partial r} - {u_{\theta} \over r^2}\right) + {v_0 \over \mu_s}\left( 1 - {\omega r \over v_0}\right) &=& 0
 \eea
In the above set of equations $E$, $\nu$ and $h$ are, respectively, the Young's modulus, Poisson's ratio, and thickness for the sheet---the connection between these values and the  parameters used in the simulations is discussed in SI Text~S1. The parameters $v_0$ and $\mu_s$ are the self-propelled speed and effective motility per unit area of the tissue. Since we presume that all cells are motile, $v_0$ is the essentially same as the self-propelled motility value used in our simulations for the tissue. The parameter $\mu_s$ is related to the motility of single cell as $\mu_s \rho =  \mu$, where $\rho$ is the cell density, or number of cells per unit area of the tissue.

The angular velocity $\omega$ is an unknown in this problem and can be obtained as follows. The equation of equilibrium for the tissue in the simplest form is:
 \bea
 \nabla.\sigma + {\rho \over \mu} (v_0  - \omega r) \hat{{\mathbf t}} = 0.
 \eea
 Taking a cross product on both sides with ${\mathbf r}$, the position vector with respect to the center, and integrating this over the entire area of the circle we get:
 \bea
{\mathbf e}_z. \int {\mathbf r} \times \nabla.\sigma dA  + \mathbf{e}_z. \int{\rho \over  \mu} (v_0  - \omega r) {\mathbf r}\times \hat{{\mathbf t}}dA = 0.
 \eea
 Since the tangential traction is zero, by design, on the boundary, the first term of this equation reduces to zero by the divergence theorem. The second term can be simplified, further, due to the presumed radial symmetry to give the following expression for $\omega$:
\bea
\omega = {4 \over 3}{v_0 \over R},
\eea
where $R$ is the radius of the confined tissue. This derivation is similar in essence to that done in Ref.~\cite{doxzen2013guidance}. Substituting this value for $\omega$ we can now solve the two equations subject to two boundary conditions:
\bea
u_r(R) = 0 \mbox{ (confinement)}, \;\;\;\; {\partial \over \partial r}\left ( {u_{\theta} \over r} \right )_R = 0 \mbox{ (shear traction)}.
\eea
Using these boundary conditions, and noting that, by symmetry $u_r(0) = u_{\theta}(0) = 0$, we get the following solution for the displacements (with respect to the undeformed configuration)
\bea
\nonumber
u_{r} &=& 0 \\
u_{\theta} &=&  {v_0 \rho (1 + \nu) R^2 \over 3 Eh \mu}\left( {r \over R} \right)^2 \left ({r \over R} - 2 \right)
\eea
The internal shear $\tau_{r\theta}$ (per unit height $h$) for the tissue is given as
\bea
\tau_{r\theta} = {Eh \over 2(1 + \nu)}\left({\partial u_\theta \over \partial r} - {u_{\theta} \over r} \right) = {v_0 R \over 3 \mu_s}\left({r \over R} \right)\left({r \over R} - 1\right)
\eea
The maximum value $\tau_{\rm max}$ of $\tau_{r \theta}$ happens at $r = R/2$ and given as
\bea
\label{tmax}
\tau_{\rm max} = {v_0 \rho R \over 12 \mu} \;\;\mbox{ and }\;\; \epsilon_{\rm max} = {v_0 \rho R (1+\nu) \over 12 \mu E h}
\eea
This simple expression, in combination with simulations,  gives us significant insights into the variables and mechanisms that influence the behavior of the coherently rotating tissue.

It can be seen from Eq.~\ref{tmax} that the maximum shear strain in the tissue, assuming it to be linear, elastic, homogeneous, is directly proportional to both $v_0$ and $\rho \sim N$. Because shear strain governs transition from solid to plastic flow~\cite{barber2010}, we expect the quantity $v_0 \times N$ to govern the transition of the tissue from solid-like to fluid-like. Since we observed solid-like coherent rotation of the tissue for $N = 140$ and $v_0 = 1$, we can expect the tissue to behave in a similar manner for $N = 170$, if the cell motile speed is taken as $v_0 = 140/170 \approx 0.82$. It can, however, be seen from Fig.~\ref{fig:4} (b), that even for values of $v_0$ as low as $0.3$, the tissue still undergoes fluidisation. Only when $v_0$ reduces to values lower than $0.1$, does the tissue recover back its solid-like behavior. This implies that there is possibly a density dependent shear threshold which controls the solid to fluid transition of the tissue. This can be achieved if confinement introduces a density dependent shear pre-strain in the tissue, thus \emph{apparently} altering the critical threshold.

The underlying mechanism of tissue fluidisation at higher densities can perhaps be understood from analysing the deformation pattern of cell triangles in the tissue. In our study, the cells, represented by their centers, are confined in a circle of a given radius $R$. The presence of confinement results in pre-straining of cell-cell connections (springs). Unlike in a homogeneous material, due to discrete nature of this system, pre-straining by circular confinement is not uniform; instead, it leads to non-uniform deformation of the springs resulting in the presence of shear pre-strain in the system, which, by definition, implies distortion of the tissue. This distortion is further enhanced by the shear strain induced by the motile forces on the cells. Delaunay triangulation, which is used in our model to obtain/update the connectivity of cells, seeks to minimise distortion (shear) in the connecting triangles that form the tissue. Everything else remaining the same, crowding increases the amount of pre-strain, and hence the initial shear strain in the tissue. Thus a crowded tissue is more susceptible to connectivity update via neighbor changes (i.e., T1 transition), which is reflected as non-zero shear strain rate or fluidisation (see SI Sections~Text~S5~and~Text~S6 for detailed analysis).

It may be noted that, in order to provide a general and more realistic, continuum description of the tissue here, one may need a more sophisticated model ~\cite{tissue2015}. This is beyond the scope of this paper, due to the difficulty in both obtaining appropriate rheology that is compatible with the discrete model, and obtaining an analytical solution. We hence, present a simple semi-analytical case of a simple Newtonian fluid to demonstrate coherent rotation for a fluidised tissue, and resort to the simulation results to make any contact with experiments (see SI~Text~S3 for derivation). The analytical predictions of mean vorticity ($\omega = {4 v_0 \over 3 R}$ in the case of solid-like and $\omega  = {v_0 \over R}$ in the case of fluid-like (calculation shown in SI~Text~S3)) closely matched the simulation values, and predict a reduction in mean vorticity with increase in cell density (Fig.~\ref{fig:3}(c)).

\subsection*{Effect of tissue size, cell stiffness and cell cohesivity on tissue fluidisation}

\begin{figure}[H]
\centerline{\includegraphics[width=11.4 cm]{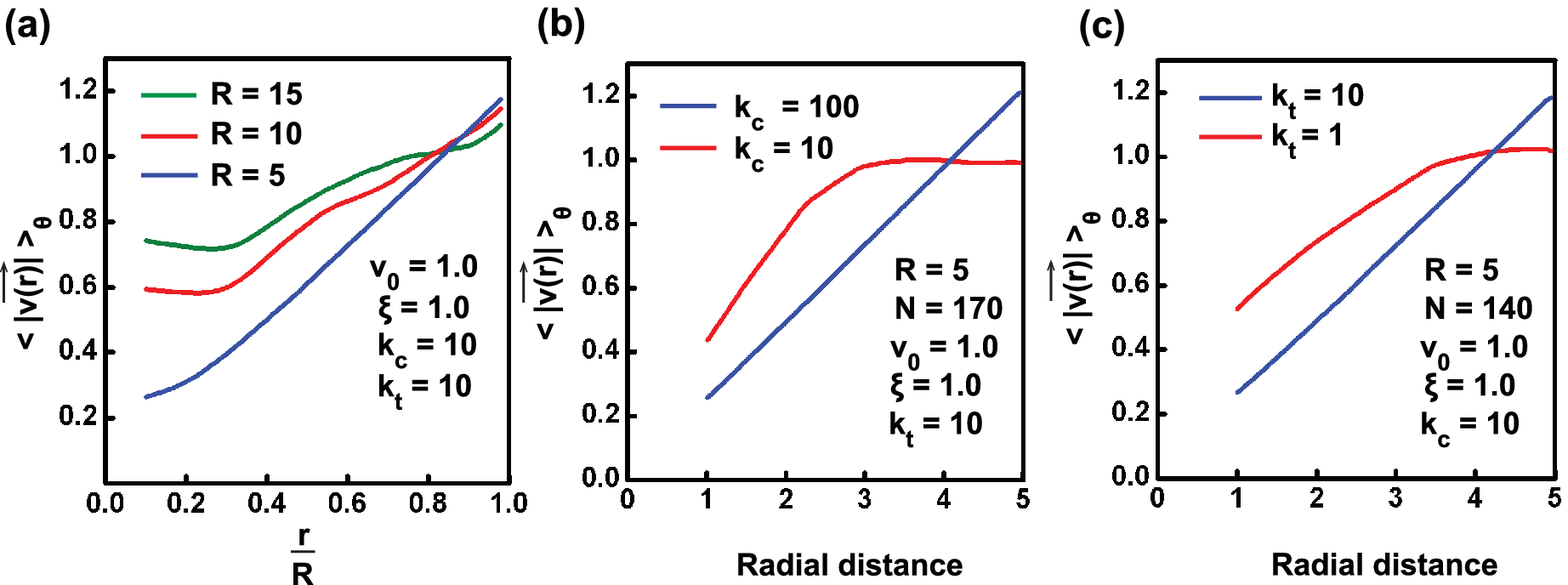}}
\caption{
\label{fig:5}
{\bf Tissue size, cell stiffness and cell cohesivity influence the fluid-like behavior of tissue. }{(a) The relationship between velocity and radial distance is examined for three systems with varying radius, while keeping the number density approximately same for all. The number of cells in the systems are taken as N = \{1170, 520, 130\} for R = \{15, 10, 5\}, respectively. The values of other parameters are chosen as that of the previous simulations. It is observed that, while keeping the number density constant, with increase in system size, the velocity versus radial-distance profile become less linear as more number of cells tend to move with a velocity comparable to $v_0$; this shows the presence of shear strain rate in the system ( see SI Video~S19). (b) Increase in cell stiffness by increasing the value of compressive stiffness ($k_c$) of a system will make the system stiff and resulting rotational behavior will be more like a solid. (c) Reduction in cell cohesivity ($k_t$) leads to fluid-like tissue behavior.}
}
\end{figure}

As seen from the previous section, in addition to providing rotational velocity, the continuum modeling also gives us a simple expression for maximum shear strain (stress) in the tissue (Eq.~ \ref{tmax}). This equation gives us further insights into the possible behavior of the tissue. For example, this expression predicts that a tissue with larger $R$ has greater shear strain, and is hence more susceptible to cross over the critical strain threshold and exhibit fluidisation.  To test this prediction, we performed simulations with increasing $R$, such that the number density of cells in the tissue was very close to the number density for the case $R = 5, N = 130$, where the tissue rotates as a solid. It can be seen from Fig.~\ref{fig:5}(a) and SI Video~S19 that, though there is no fluidisation for $R = 5$, for larger $R$, the tissue behaves in an increasing fluid-like manner---more and more layers of tissue were observed to move with velocity close to $v_0 = 1$.   Thus, tissue can undergo fluidisation solely due to the influence of system size. The relatively larger values of cell speeds at lower radial distance is due to radial movement of cells (see SI Video~S19), and is possibly related to the dominance of radial modes with increasing system size (SI Section~Text~S2). Thus, even though Eq.~\ref{tmax}, does not exactly capture the tissue behavior with increasing system size, it provides us with pointers in the right direction, and concurrently exposes the shortcoming of describing the tissue as a solid-like material~\cite{doxzen2013guidance}.

It can be noted from Eq.~\ref{tmax} that, the shear strain is, as expected, inversely proportional to the tissue stiffness. This implies that tissues with stiffer cells ($k_c$) and greater cell-cell cohesivity ($k_t$) are less susceptible to cross over the critical strain threshold and more likely to exhibit solid-like behavior; the inverse would apply for tissues with softer cells. For the case $R = 5, N = 170$, increasing the stiffness $k_c$ for a tissue  from $10$ to  $100$ results in a transition from fluid-like to solid-like coherent rotation of the  tissue (Fig.~\ref{fig:5}(b)). Similarly decreasing the value of cell cohesivity ($k_t$) also leads to fluid-like behavior of tissue. We can see from Fig.~\ref{fig:5} (c) that for $N = 140$ when $k_t = k_c = 10$, then the velocity profile being linear is an indication of rigid body rotation (as per the analytical solution for elastic solids shown in previous section). However,  when $k_t$ is decreased from $10$ to $1$ while keeping $k_c = 10$, then it is clearly seen that the tangential velocity as a function of radial position has saturating profile (as seen in the previous section for analytical solution for viscous fluid) indicating fluidisation. Thus the stiffness and cohesivity of  tissue cells can independently control the nature of coherent rotation for the confined tissue.

\subsection*{Coherent rotation in non-convex annular geometries}
While the above results of coherent rotation were obtained in circular geometries, it remains unclear if similar coherent rotation is also possible in non-convex geometries. Of the various non-convex geometries, annular rings are often observed \textit{in vivo} in glands, ducts or tissues with lumen inside. Several studies have probed the collective behavior of cells inside annular geometries~\cite{rolli2012, nelson2005emergent, li2014coherent}. Since annular geometry represents the simplest non-convex geometry obtained from a circular shape, we next studied the coherence patterns in annular geometries and the influence of cell density. For this, simulations are done with outer and inner radius of annulus taken as $100~\mu m$ and $70~\mu m$. Simulation with $N=100$, $k_c=10$, $k_t=10$, $\xi=1$, $v_0=1$, shows that here also, after a short initial transient mode, cells exhibit robust coherent rotation similar to that on circular geometries (Video S12). However, the pattern of coherent rotation is dictated by the stiffness of cell-cell adhesions. Specifically, for lower stiffness values ($k_t$), different cell layers in annular section may move in different directions (Video S13); in contrast, for higher stiffness values of cell-cell connections, cells move in a robust manner after an initial breathing mode (Video S12).

\begin{figure}[H]
\centering
\includegraphics[width=11.4 cm]{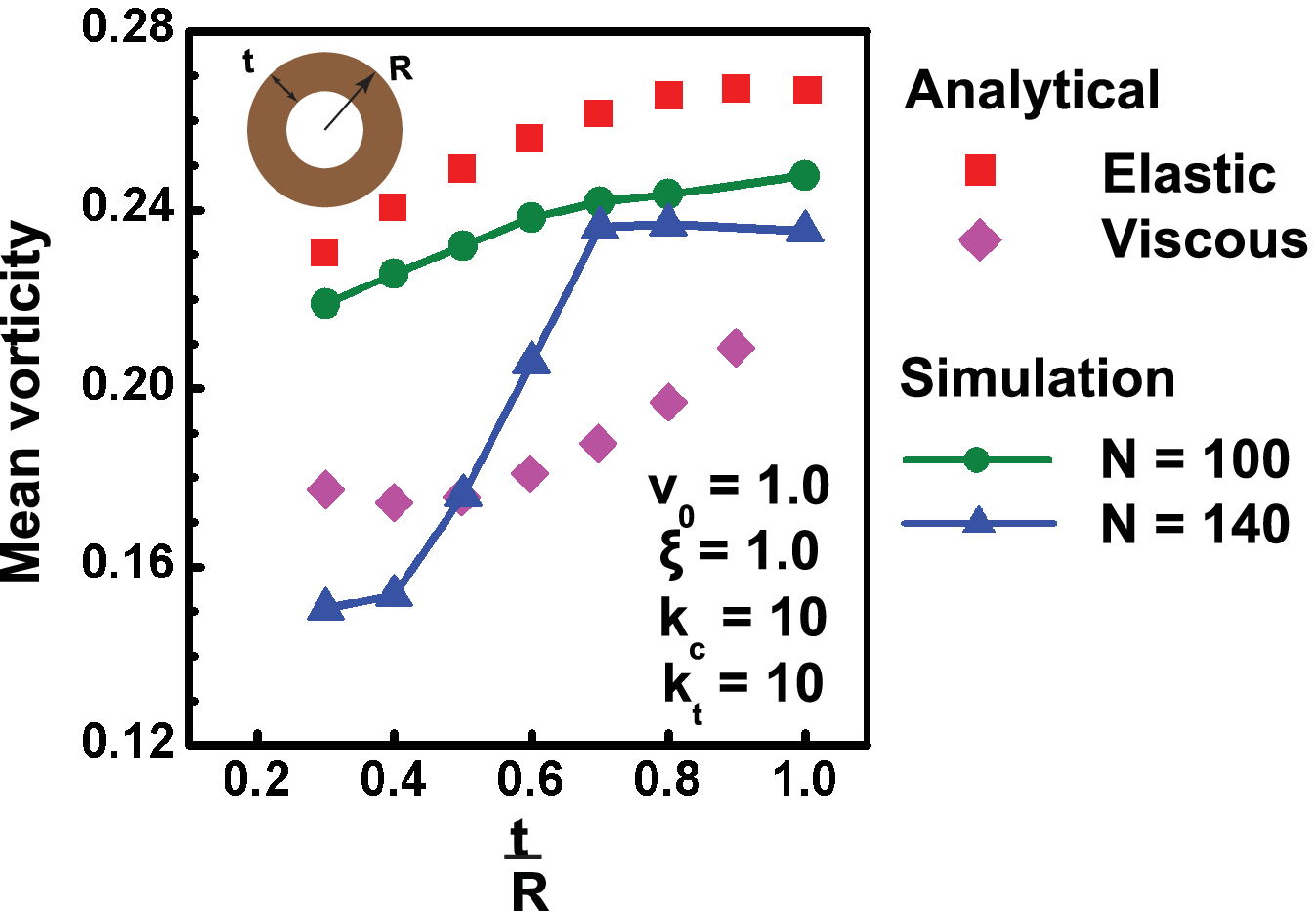}
\caption{\label{fig:7} {\bf Coherent rotation of cells confined in annular geometry}. As observed for circular substrates, cells confined inside annular geometries also exhibit coherent rotation. Simulations done on an annular shaped geometry with outer radius, $R$ and thickness, $t$ show that mean vorticity of system decreases with increase in number density as in the case of a circle. Furthermore, simulations done with two sets of cell numbers (N=100, represented by green curve and N =140, represented by blue curve) show that at lower densities, system behaves like an elastic solid, roughly matching their values with analytical results (red points). For higher number of cells, cell behavior is more like an elastic solid for thicker sections. As the thickness of annulus reduced, cell state transitioned from an elastic solid to viscous fluid. The red and magenta curves showing the analytical values of elastic solid and viscous fluid respectively, are derived as explained in main text.
}
\end{figure}

Next, to test the effect of cell density on mean vorticity, simulations were performed on annular geometries with varying annular thickness, $t$ and constant outer radius ($R$). For a constant number of cells ($N$), varying $\frac{t}{R}$ leads to change in number density, and is hence expected to influence the pattern of coherent motion.
Consistent with this, distinct behavior is observed for two different values of $N$. Figure~\ref{fig:7} shows the plot of mean vorticity of system as a function of $\frac{t}{R}$ for $N = 100$ (green curve) and $N = 140$ (blue curve); the other parameters are kept the same as in previous simulations. As seen from the plot, it is seen that with increase in the thickness of annulus, the mean vorticity of the system also increases, which matches with the findings of Li and Sun~\cite{li2014coherent}. In addition to that, we have also shown that for lower number of cells ($N=100$), system behaves more like an elastic solid with minimum shearing between cell layers similar to that seen in the circular geometries. But it is interesting to note that for larger number of cells ($N=140$), system behaves like an elastic solid at higher $\frac{t}{R}$ values. However, when the thickness of annular section decreases, cells become more compressed which leads to the fluidisation of system and as a result, for lower $\frac{t}{R}$ values, system behavior is more similar to viscous fluid. In order to have a better understanding, we have also calculated the analytical values of mean vorticity of system using the following equation:
\begin{equation}
\omega = {1 \over 2}{2 \pi (R_{\rm out} v_{\rm out} - R_{\rm in} v_{\rm in}) \over \pi (R_{\rm out}^2 - R_{\rm in}^2)},
\end{equation}
from Stokes theorem~\cite{chadwick2012}. Here $R_{\rm out}$ and $R_{\rm in}$ are the outer and inner radius and $v_{\rm out}$ and $v_{\rm in}$ are the outer and inner velocities, respectively. As seen from
Fig.~\ref{fig:7}, the analytical values of vorticity (see earlier sections for analytical expressions for elastic and viscous
calculation for $v_{\rm out}$ and $v_{\rm in}$) closely follow the computed values and illustrate the dependence of vorticity on $\frac{t}{R}$ values. Taken together, our findings on circular as well as annular geometry imply that for different confinement geometries, cell behavior can vary between that of a perfectly elastic solid and a complex fluid depending on cell density.

\subsection*{Sensitivity of coherent rotation to cell division}
In addition to illustrating the role of confinement in inducing coherent motion, our results also demonstrate the critical influence of cell density in dictating the pattern of coherent motion. While cell density can be experimentally controlled in \textit{in vitro} experiments, under \textit{in vivo} conditions, cell density is controlled by cell division---a factor that was not taken into account in our simulations. While several computational studies have tried to understand how cell division influences morphogenesis~\cite{kraeussling2011highly, priori1996mathematical, basse2003mathematical}, the sensitivity of coherent motion to cell division remains unexplored. Having demonstrated the robust influence of cell density in our simulations, we next probed the extent to which coherent motion is sensitive to changes in cell density effected by cell division events. Cell division can occur either synchronously (i.e., all cells divide at the same time) or asynchronously (i.e., cells divide at different times). During early stages of embryo development, cells generally exhibit multiple fast synchronized division, accompanied by a transition stage and subsequent slow non-synchronized divisions, with different cells having different stages of cell cycle~\cite{newport1982major, kane1993zebrafish}. {\it In vitro} the cell cycle of individual cells can be synchronized by serum starvation~\cite{kato1992synchronous}. To study the sensitivity of coherent motion to cell division, we probed how changes in the total number of cells in a confined geometry would influence the pattern of rotation. For these studies, an annular geometry was chosen, as such geometries are biologically relevant\cite{friedl2009collective}. Further, both synchronized and asynchronized cell division were introduced into an already rotating system to perturb the steady state rotational motion.

\begin{figure}[H]
\centering
\includegraphics[width=11.4 cm]{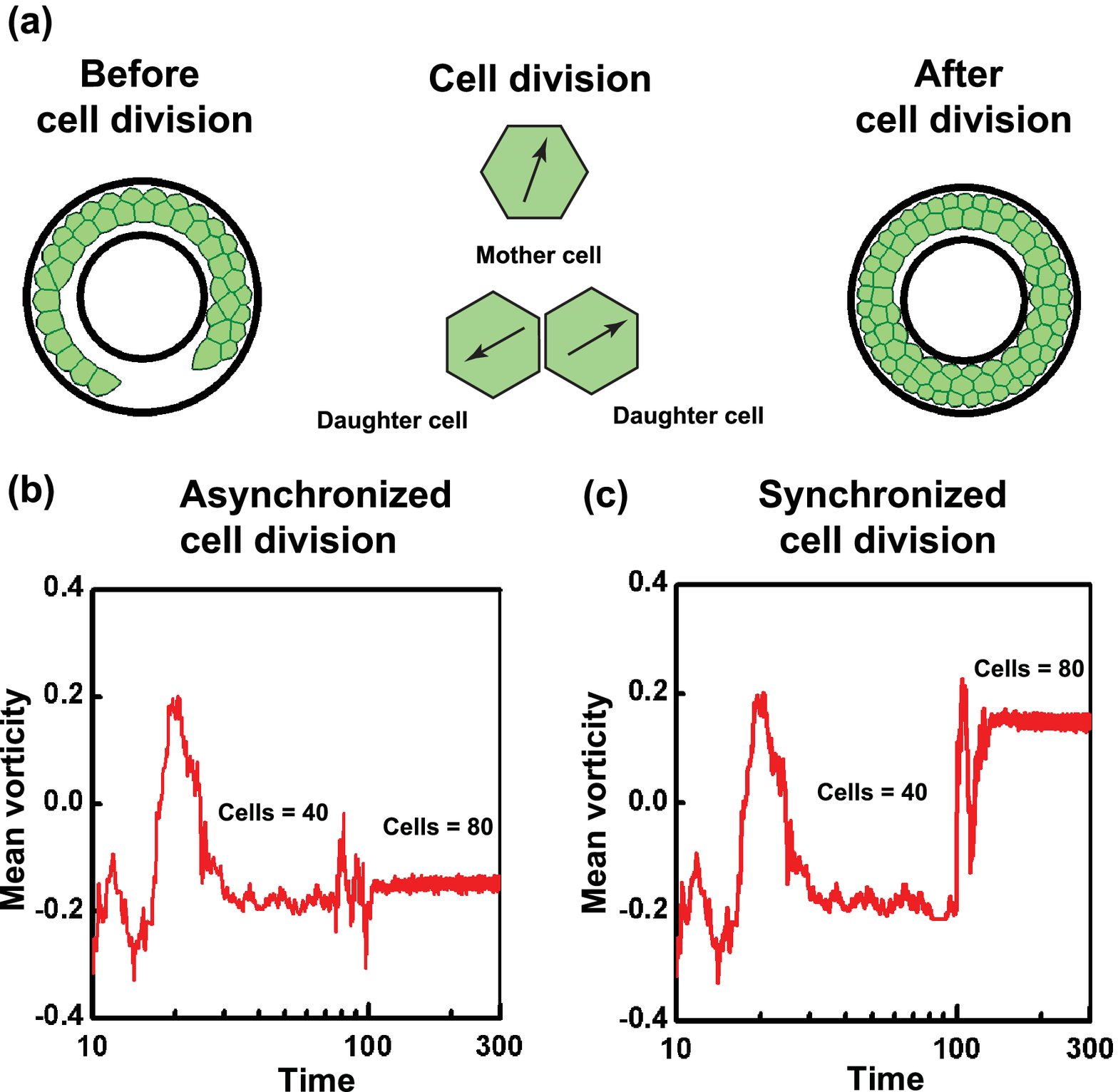}
\caption{\label{fig:8} {\bf Synchronous cell division changes the sense of coherent rotation}. (a) Cells are allowed to undergo division, where each mother cell on attaining a mature cell cycle number will divide and become two daughter cells with equal and opposite polarization (as indicated by black arrows inside the daughter cells). Two cases are analysed where cells are allowed to divide either synchronously or asynchronously. For both cases, initially starting with 40 number of cells, division is continued till number of cells become 80. (b) Even though incapable of change the direction of overall rotation, asynchronized cell division causes local perturbations in the pattern of rotation, which shortly dies down and system continues the steady rotational mode, indicated by the same sign of mean vorticity before and after cell division. (c) In the case of synchronous division of cells, the large perturbations introduced into the system in a short time is capable of inducing the change in the direction of rotation indicated by the opposite signs of mean vorticity before and after the division. Even though the reversal in the direction of rotation after synchronous  division has not happened in all the cases analysed, we observed a preferential bias in the change in direction tendency.
}
\end{figure}

A total number of 40 cells, which are below the level of confluence, were confined in an annular substrate of outer radius 100~{$\mathrm{\mu m}$} and inner radius of 70~{$\mathrm{\mu m}$}, and allowed to reach a state of coherent rotation (Fig.~\ref{fig:8}(a)). Once this state was reached, cells were allowed to divide either synchronously or asynchronously. For implementing asynchronous cell division, each cell in the system was initially assigned a random cell cycle number between 0 and 1. In contrast, for synchronized division, the initial cell cycle number of all cells were set to $0$. The time interval for cell division was assumed as 24 hrs for all our simulations, and represents the time taken by any cell to reach its cell cycle from 0 to 1. Any cell, on reaching a cell cycle number of 1, underwent division to form two new daughter cells, provided the cell area was above some critical area (see Materials and Methods). The two new daughter cells formed after division, were assigned equal and opposite polarization in random direction, and placed at $a_0/2$ along the major principal axis of mother cell's area (Fig. \ref{fig:8}(a)). For both synchronous and asynchronous division, cell division was stopped when the total cell number reached 80.

Interestingly, synchronous and asynchronous division were found to perturb coherent rotation to varying extents.
Asynchronous division did not alter the direction of rotation, but only created some local disturbances, after which coherent rotation was fully established. This is clearly seen from the temporal profile of the mean vorticity ( Figs. \ref{fig:8}(a, b)) where transient fluctuations in the mean vorticity quickly die down and cells continue to rotate coherently. In contrast, for the case of synchronous cell division, on several occasions, the direction of coherent rotation underwent a change as observed from the change in mean vorticity values
(Fig. \ref{fig:8}(c)). Statistical analysis revealed that out of 200 independent simulations conducted, this reversal after synchronous cell division was observed for almost 60\% of the cases, indicative of a preferential bias for the change in rotation (Video S14--S15). Together, these results suggest that while coherent rotation is insensitive to asynchronous division, synchronous division introduces a bias in the direction of rotation. However, the biological implication of this reversal remains to be established.

\subsection*{Effect of removal of confinement: cell stiffness and cell-cell cohesivity dictates invasion pattern from coherent motion}
\begin{figure}[H]
\centering
\includegraphics[width=11.4 cm]{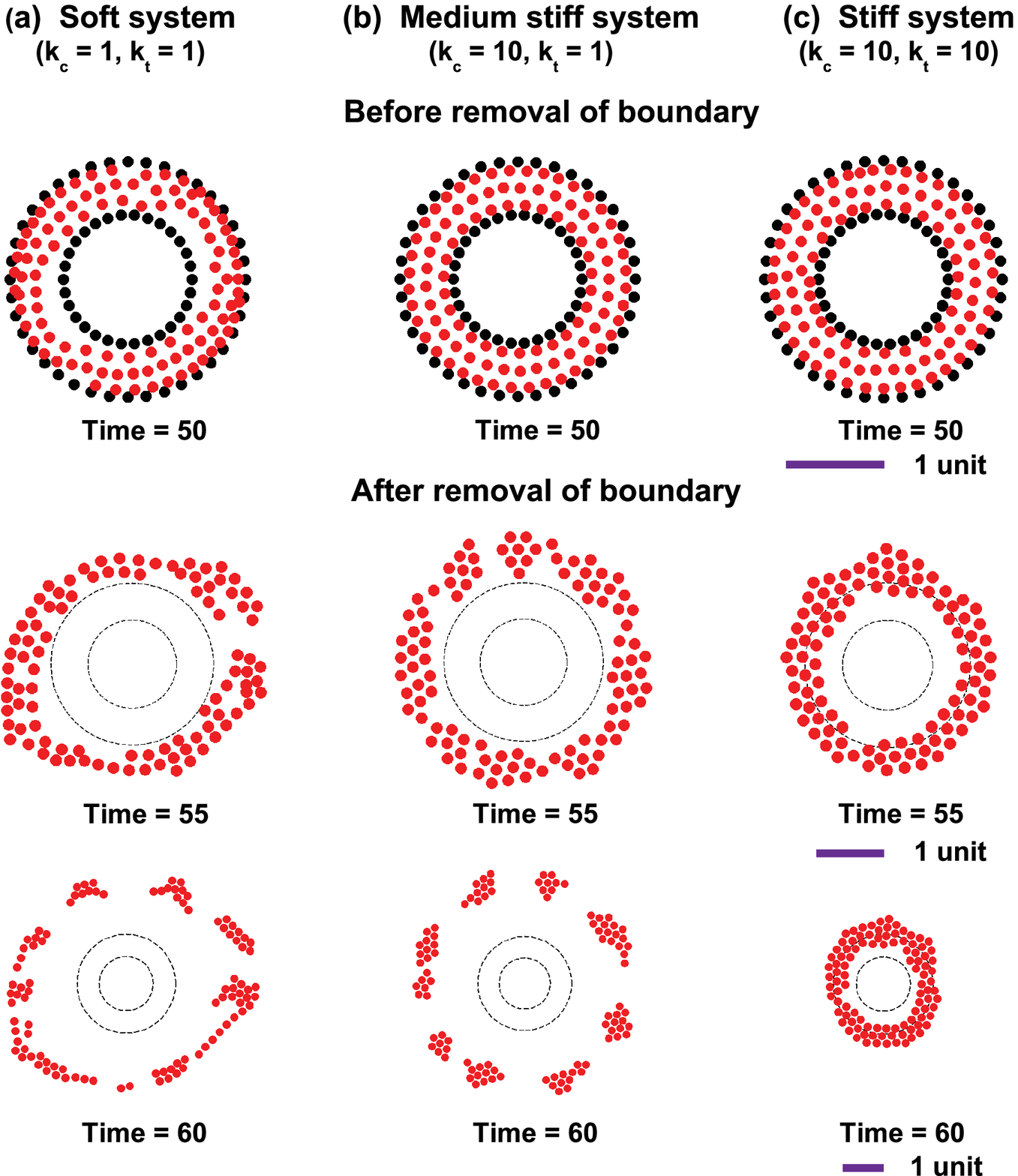}
\caption{\label{fig:9} {\bf Cell stiffness and cohesivity dictate invasion pattern from coherent motion.}
Three different systems of cells are taken with different stiffness of
cell-cell connections. Simulations for (a) a soft system with $k_c=1$ and $k_t=1$
(b) a medium stiff system with $k_c=10$ and $k_t=1$ (c)  stiff system with $k_c=10$ and $k_t=10$. The number of cells in all the three cases are same and equal to 100. After reaching a steady state of rotation,
confinement was removed at time, $t= 50$. The snapshots of cell migratory
patterns at $t=55$ and $t=60$ are also shown. For the case of intermediate stiff system, cells migrate in clusters compared to softer system where cell invasion pattern is more scattered. At the highest stiffness, cells continue to rotate even after removal of boundary. The length scale for each set of figure is shown below them.}
\end{figure}

Under \textit{in vivo} conditions, the confinement assumed in our simulations, is generally provided by the surrounding extracellular matrix (ECM). For example, all epithelial tissues are surrounded by the basement membrane, which helps to maintain tissue organization and prevents cell invasion. However, the basement membrane is breached by epithelial cells which turn cancerous. Cancer cells are known to invade both as single cells and collectively~\cite{friedl2003tumour, haeger2014cell, czirok2013collective}. Since coherent rotation is sensitive to the properties of cell-cell contacts (i.e., $k_t$ and $k_c$, respectively) (Fig.~\ref{fig:5}), we hypothesize that, the initial coherent rotation dictated by the properties of cell-cell adhesions has a distinct bearing on the eventual invasion pattern, when confinement is removed. To test this hypothesis, we have studied the invasion patterns formed when a coherently moving group of cells break their boundaries and invade to the surrounding matrix. For doing this, three conditions were chosen with the following combinations of $k_t$ and $k_c$ to mimic different properties of cells and cell-cell adhesions: $k_c = k_t = 1$ (i.e., soft), $k_c = 10, k_t = 1$ (i.e., medium stiff), and $k_c = 10, k_t = 10$ (i.e., stiff). The number of cells in each system was taken as 100 and the values of all other parameters were kept the same as that of other simulations. Once coherent rotation was set up in all the systems, the confinement was relaxed at $t=50$ to allow for invasion. Consistent with our hypothesis, the combination of $k_c$ and $k_t$ were found to directly influence the nature of coherent motion (Fig. \ref{fig:9}(a--c) and Video S16--S18). For the soft and medium stiff systems, the extent of invasion (i.e., radial position as function of time) remained the same. However, contrary to the `soft' case where cells scatter in all directions, for the `medium stiff' case, cells move radially outward as clusters which remain connected. For the `stiff' case, cells continue to rotate even after the removal of confinement. Together, these results demonstrate that the nature of coherent motion set by the extent of cell-cell cohesivity dictates the invasion pattern when confinement is removed. Also, the persistent rotation of stiff cells with stiff adhesions even after the removal of boundary shows that even though confinement is essential for the emergence of coherent rotation, depending upon the properties of the system, the presence of a confinement is not mandatory condition for the cells to continue in their coherent motion.

\section*{Discussion}
Coherent rotation of cells is essential for a variety of physiological activities. \textit{In vitro}  studies have reported the occurrence of robust rotation of epithelial sheets under confinement. We developed a self propelled, cell center-based model to understand this phenomena. We have shown both numerically and analytically that, a two-way feedback between cell velocity and cell polarization, along with force transmission via cell-cell connections, is sufficient to produce persistent rotational mode of migration for cells under confined conditions. Though there are computational studies that demonstrate the possibility of CAM for tissues in confined geometries, we provide a logical and plausible explanation as to why this may happen. Similarly, we show the presence of additional hydrodynamic modes that deform the tissue in the radial direction, if the time-scale of relaxation of these modes is comparable to the time-scale of the orientation of polarization. We also predict that such radial modes are more likely to be observed for larger size tissues since they can sustain low wavelength and slow decaying radial modes, whose stiffness is inversely proportional to the tissue size---this finding is consistent with the experimental observations of Deforet et al.~\cite{deforet2014}.

Our model predicts that irrespective of the size of the confinement, the cells can reach the state of coherent rotation. This implies that the velocity correlation length for the tissue can be as large as the system size. However, Doxzen et al. reported absence of coherent rotation for larger confinement size and  linked this observation with the system size being larger than the experimentally observed correlation length of $\approx 10$ cell lengths that was reported elsewhere~\cite{petitjean2010velocity}.  This apparent contradiction regarding correlation length can possibly be explained as follows. Unlike confined tissues, the reported correlation length were obtained for systems having free boundaries. It is known that the presence of free boundaries leads to modifications in the tissue boundary conditions (e.g., leader cells, high cable tension, etc.)~\cite{tarle2015}. These modifications may influence the velocity correlation lengths for the tissue, and therefore lead to qualitatively different behaviors when compared with that of confined tissues. We would also like to point out that for the same type of cells, velocity correlation length can be influenced by mechanical perturbations---time-increasing velocity correlation length of up to $450~\mu$m was reported in Ref.~\cite{angelini2010} for migration cells on deformable substrates. Thus the correlation length need not be an inherent property of a tissue type, and can be influenced by mechanical perturbations (such as confinement). In addition to these factors, perturbations in the form of cell division and cell death, can also influence the temporal dynamics of velocity correlation length of a confined tissue, and may be one of the reasons why coherent rotations were experimentally not observed for larger size tissues.

Our results also demonstrate that a few experimentally controllable/observable variables---cell density ($N$), motile speed ($v_0$), system size ($R$), cell stiffness ($k_c$) and cell cohesivity ($k_t$)---collectively tune the behavior of a coherently rotating tissue between that of an elastic solid and a complex fluid. We found that, in our model, tissue fluidisation is associated with increased shear strain in the system, which can be relieved by connectivity changes (T1 transitions). Such T1 transitions that are dependent on shear deformation of cells, have been also observed/modeled in a very recent work by Etournay et al.~\cite{etournay2015},  thus providing biological relevance for our modeling.  We found that upon increase in cell density the tissue showed an increasingly fluid-like behavior. This effect is not expected if the apparently radially symmetric confinement isotropically compresses the tissue. However, due to the discrete nature of the tissue, we observed that for greater cell density, the confinement induces larger distortion within the cells and makes the tissue more susceptible to fluidisation. Moreover, we also found that other factors such as  increase in cell motility, increase in system size, and decrease in cell stiffness, which lead to increase in shear strains in the tissue, can make the tissue more vulnerable to fluidisation. Thus, our study identifies some of the potential parameters, which, as described above, could give rise to novel and hitherto unreported behavior for confined tissues.

The basement membrane, which is found at the basal surface of epithelial cells is essential for tissue polarity, and maintains tissue structure by confining cells. In epithelial cancers, uncontrolled proliferation of cells leads to buildup of stress within the tissue. Subsequently, malignant cells breach the basement membrane and escape into the surrounding stroma. Cancer invasion through these matrices is dictated both by extrinsic factors (e.g., ECM density and organization) and by intrinsic factors. Among the intrinsic factors, our findings implicate cell division and functional nature of cell-cell contacts as two parameters influencing the coherent motion, and the invasion process. Our studies show that synchronous division introduces a bias in the direction of rotation, with a reversal in the direction of rotation observed in nearly 60\% cases. Whether or not this reversal has any significance in invasion remains to be established.

EMT, or epithelial to mesenchymal transition, refers to the complex process whereby immobile epithelial cells lose their cell-cell adhesions and get converted into motile mesenchymal cells \cite{nieto2011ins,lamouille2014molecular}. EMT is relevant both to normal embryonic development and in carcinogenesis. During EMT, downregulation of the cell-cell adhesion protein E-cadherin is accompanied by upregulation of mesenchymal cadherins like N-cadherin which favour forming of transient contacts \cite{peinado2004transcriptional}. However, EMT is not an all-or-nothing phenomenon and cells can exist in partial EMT states. Such states have been reported in carcinosarcomas \cite{sarrio2008epithelial}. In contrast to EMT, cancer cells are also known to exhibit collective cell migration, where E-cadherin-positive cell-cell contacts are maintained \cite{haeger2015collective}. It is likely that both cell scattering and collective cell invasion are outcomes of alterations in the physical behavior of cell-cell contacts. This was evident from the scattering patterns observed in our simulations when confinement was removed. When adhesions were soft, cells scattered in all directions with the formation and breakage of transient adhesions. This mode of invasion was closer to that of single cell invasion. However, when adhesions were medium stiff, cells scattered as uniform-sized clusters indicative of a collective mode of invasion. Our results thus suggest that the different modes of invasion observed in different contexts are dictated by the strength of cell-cell adhesions. When adhesions were very strong, even upon removal of confinement, the cell layer expanded to release the confinement-induced compression, but continued to exhibit coherent rotation. Under this condition, activation of invasion as observed experimentally after removal of confinement \cite{rolli2012}, is likely to involve mechano-chemical changes in the motility behavior of cells arising from the presence of free boundary.

In conclusion, our framework of velocity-polarization coupling successfully recapitulates coherent motion in confined circular and annular geometries, demonstrates the influence of  a few experimentally controllable variables---motile speed ($v_0$), cell density ($N$), cell stiffness ($k$) and system size ($R$)-- in collectively dictating the pattern of coherent motion, and illustrates the effect of synchronous cell division on coherent motion. In addition, our model predicts the invasion patterns that arise due to coherent motion when confinement in removed. Future work can be focused on further improving the predictive power of our model by incorporating the effect of actomyosin contractility and substrate properties (e.g., stiffness).

\section*{Materials and Methods}
\subsection*{Methodology}
For simulations, the scaling quantities for length, time and force are taken as $a_0$, $a_0/v_0$ and $v_0/\mu$ respectively. Unless otherwise specified, values of $a_0$, $v_0$ and $\mu$ are taken as $1$ for all the simulations. The cells were represented by their centers, and their connectivity was obtained via Delaunay triangulation~\cite{pathmanathan2009computational}, which produces least number of distorted triangles, i.e., triangles with least shear strain~\cite{delaunay}.
Delaunay triangulations are dual to Voronoi tessellations (see Fig.~\ref{fig:1}(b)) and the Voronoi polygon for a given cell center can be modeled to represent the cell~\cite{pathmanathan2009computational, mirams2013chaste}. It is unrealistic to obtain the areas of boundary cells directly from tessellation, since the Voronoi polygon for these cells can have a vertex at infinity~\cite{delaunay}. To circumvent this problem, a row of dummy points are inserted at the boundaries, just to create well defined polygons for visualization, but they do not contribute to the dynamics of the system. Even though cells are connected to each other by cell-cell connections to form an apparently solid tissue, based on the dynamic position of the cells, this connectivity is constantly updated using Delaunay triangulation and may result in cell neighbor changes within the tissue. These modification of neighbors can be interpreted as the so called T1 transitions, the specialised terminology for neighbor exchange in the context of foams and epithelia~\cite{fletcher2014vertex} (also see SI Section VI).  Since confinement is experimentally shown to be essential for setting up coherent rotation, we model the soft confinement at boundaries by providing resistance of stiffness $3 k_c$ at the edges, which will apply force on any cell trying to cross the boundary and thus prevent them from escaping.

To begin with, cell centers were randomly  distributed inside a confined zone of given dimensions, and were allowed to equilibrate, such that the velocities of all the cells was near zero---the cell-cell connectivity was obtained by Delaunay triangulation, as described.
Since we model cells as self propelled particles, they were assigned a uniform motility, $v_0$ in random directions of their polarization after the equilibration stage. This led to the evolution of position and polarization of cells thereby setting up of the dynamics of the system. After a short initial transient state with random motility, cells started to rotate coherently and reached a steady state of motion. However, it is to be noted that the current formulation does not account for the effect of any noise in the system. For solving the set of differential equations numerically, we adopted forward Euler scheme that was implemented in Matlab. After performing a detailed convergence study, a time-step of $\Delta t = 0.001$ was used for all the calculations. In order to quantify the angular motion of tissues, we calculated the mean vorticity of system derived from the antisymmetric part of velocity gradient matrix. Mean vorticity can be defined as $\frac{\int\omega dA}{\int dA}$, where $\omega=\frac{1}{2}\left(\frac{\partial u}{\partial y}-\frac{\partial v}{\partial x}\right)$ is the vorticity tensor, and $u$ and $v$ represent the velocity components in $x$ and $y$ directions at the time $t$.

As already described, polarization of cells are initially randomly oriented. So it is logical to assume that there should not be any preferential bias in the direction of coherent rotation of cells. In order to verify this, a statistical analysis is carried out. Out of 100 independent simulations performed on both circular and annular geometry without cell division, almost equal number of clockwise and counter-clockwise rotations are obtained, which shows that there is no preferential bias in the system. Similarly, 100 independent simulations performed in an annular geometry with asynchronized cell division show no change in their direction of rotation after cell division. In order to check the statistical significance of the switch in the rotational direction on synchronous cell division, two sets of 200 independent simulations are carried out. For the first set of simulations, the polarization of daughter cells are assigned equal and opposite in random direction, while for the second case, their polarizations are completely random and independent. For both these cases, cells are observed to preferentially (around 60\%) switch their direction of rotation after synchronous cell division, which indicate that the mechanical perturbations caused because of cell division may be the reason for the additional $10\%$ bias in switch in direction.

\setcounter{equation}{0}
\setcounter{figure}{0}
\setcounter{table}{0}

\renewcommand{\thefigure}{S\arabic{figure}}
\renewcommand{\thetable}{S\arabic{table}}
\renewcommand{\theequation}{S\arabic{equation}}
\renewcommand{\thesubsection}{S\arabic{subsection}}

\section*{Supporting Information}
 \subsection*{Text S1 --
Supplemental text showing the derivation of relation between spring
constant ($k$) and Young's modulus ($E$) for
triangular network of springs.}

Let us consider a triangular network of springs with A, B, C as the initial position of
cells and $a_0$ as the length of springs at equilibrium as shown in
Fig.~S1. Now, each cell is given
a uniform stretch $\delta$ so that their position has changed to A', B' and C'
respectively. For a network with six fold symmetry, the potential energy change associated with
this stretching can be written as
$\Delta F = 3\times \frac{1}{2} k\delta^2$\cite{boal2012mechanics}, where $k$
is the spring constant. Assuming the
system as elastic, isotropic and homogenous, the potential energy stored
in the system as the result of deformation is equal to
the strain energy of the system which can be written as
$U=\frac{1}{2} \sigma_{ij} \epsilon_{ij}V$, where $\sigma_{ij}$
is the stress tensor, $\epsilon_{ij}$ is the strain tensor
and $V$ is the volume of the element. Assuming the system as elastic and isotropic,
the stress strain relationship can be written as
$\epsilon_{ij} = \frac{(1+\nu)}{E}\sigma_{ij}-\frac{\nu}{E}\sigma_{kk}\delta_{ij}$,
where $E$ denotes the Young's modulus, $\nu$ is the poison's ratio
and $\delta_{ij}$ is the unit tensor.
For an isotropic stretch for a 2-D system, the normal strain components
will be equal and
shear will be zero. So $\epsilon_{xx}
= \epsilon_{yy} = \epsilon
= \frac{\delta}{a_0}$ which will give the stress components as
$\sigma_{xx} = \sigma_{yy} = \sigma = \frac{E}{(1-\nu)}\frac{\delta}{a_0}$.
Now the total strain energy of the system will be
$U = \frac{E}{(1-\nu)}(\frac{\delta}{a_0})^2 V$.
Writing volume $V$ in terms of
the height of the element, $h$ and equilibrium length of spring
$a_0$ as $V= \sqrt(3)/4\times  a_0^2h$,
and equating strain energy with the change in potential energy, the
spring constant will be
$k = \frac{Eh}{2 \sqrt 3 (1-\nu)}$

\subsection*{Text S2 --
Supplemental text showing the details of why coherent rotational motion is seen for a self-propelled elastic solid when the polarization
vector for a cell has a tendency to align with the velocity of the cell.}

The equation of evolution for cell position and polarization, respectively, for the current system are (also see main text):
\begin{eqnarray}
\nonumber
\frac{d\bm r_i}{dt}&=&v_0\hat{\bm p_i}+\mu \bm F_i,\\ \label{eq:evo}
\frac {d \hat {\bm p}_i}{dt}&=&\xi(\hat {\bm p}_i \times \hat {\bm v}_i . \bm e_z)\hat {\bm p}_\perp
\label{eq:polrule}
\end{eqnarray}
As per this evolution rule, the polarization of the cell has a tendency to align with its velocity. Additionally, the polarization direction also feeds into the velocity of the cell and tends to modify its speed and direction. Now, to understand, at least semi-analytically, the origin of the rotational motion under confinement for an elastic solid formed of self-propelling particles as given above, we use the following procedure motivated from the arguments in Refs.~\cite{ferrante2013collective,henkes2011}.  We  extend this reasoning to argue that even if the cells can exchange their neighbors, coherent rotational motion of the cell sheet is the most likely outcome.

The elastic system of springs under circular confinement is free to undergo rigid body rotation about its center. Its translational degrees of freedom are, however, curtailed due to the confinement. Using ideas from structural mechanics, let us denote the stiffness matrix of the system in its rest (or stress-free) conformation as $[K]$~\cite{leet2002}. The matrix $[K]$, is symmetric, positive semi-definite, and has dimensions of $2N_{\rm cells} \times 2N_{\rm cells}$. If the displacement of the cells (nodes) from their rest position, in terms of column matrix of dimension $2N_{\rm cells} \times 1$,  is $\{ u\}$, then the equation of motion for the system in the complete matrix form can be written as:
\bea
\label{eq:evol}
{d \{u\} \over dt} = v_0 \{p\} - \mu [K]\{u \}.
\eea
Additionally, in this case, the polarization of each cell prefers to align with its velocity (see Eq.~\ref{eq:evo}). Since the stiffness matrix $[K]$ is symmetric and positive semi-definite, it has  $2N_{\rm cells}$ orthogonal eigenvectors $\{\phi_i\}$ and corresponding non-negative eigenvalues $\lambda_i$. The only zero eigenvalue is the one corresponding to the rigid body rotation mode $\{\phi_0\}$; all other eigenvalues are positive. The displacement $\{u \}$ and velocity $\{\dot{u}\}$ can then re-written in the form of eigen-modes as
\bea
\{u\} &=& \sum_{i=0}^{2 N_{\rm cells}-1}\alpha_i \{\phi_i\}, \mbox{ and }\\
\{\dot{u}\} &=& \sum_{i=0}^{2 N_{\rm cells}-1}\dot{\alpha}_i \{\phi_i\},
 \eea
 where $\{p\}$ is the polarization of all the cells, combined in the form of a column vector of size $2N_{\rm cells} \times 1$. Expressing Eq.~\ref{eq:evol} using the eigenmodes,  we get the following set of equations in terms of eigenmode amplitudes
 \bea
 \label{eq:alpha}
 {d  \alpha_j \over d t} = v_0 \langle \phi_j \rangle \{p\} - \mu \lambda_j \alpha_j,
 \eea
where $\langle \phi_i \rangle$ is the eigenvector written in row format.  This equation can be re-written as:
 \bea
 \label{eq:alpha2}
 {d  \alpha_j \over d t} = v_0 \beta_j - \mu \lambda_j \alpha_j,
 \eea
where $\beta_j =  \langle \phi_j \rangle \{p\} $.
It seems safe to presume that for a certain time interval $\tau \sim 1/\xi$, where $\xi$ is the response rate for polarization (see Eq.~\ref{eq:evo}), the polarization remains almost constant. In this case Eq.~\ref{eq:alpha} can be solved to provide us the following solution for the amplitude $\alpha_j$ of any mode $j$.
 \bea
 \alpha_j = {v_0 \beta_j\over \lambda_j \mu} [1 - \exp(-\mu \lambda_j t)],
 \eea
 and the corresponding \emph{velocity} of the mode is given by
 \bea
 \label{eq:alphadot}
 \dot{\alpha}_j = {v_0 \beta_j}\exp(-\mu \lambda_j t),
 \eea
where we have assumed zero initial conditions for $\alpha_j$. It is very clear from Eq.~\ref{eq:alphadot} that modes with larger $\lambda$, i.e., greater stiffness or small wavelength, would decay faster as compared with the modes with smaller $\lambda$. The smallest $\lambda$ possible for the current system is $\lambda_0 = 0$, corresponding to the mode that involves rigid body rotation of the tissue. This implies that the amplitude $\alpha_0$ corresponding to pure rotation, and more importantly the angular speed $\dot{\alpha}_0$ ($v_0 \beta_0$) increases with time. The motility parameter $\mu$ sets the rate at which the energy is dissipated from mode $j \ne 0$, whereas the polarization orientation constant $\xi$ sets the rate at which the energy is pumped into each mode (in the form of $\beta_j$).

Let us first look at the case with \emph {medium} value of $\xi \approx 1$, i.e., $\xi \approx \mu \lambda_1$. The polarization column vector $\{p\}$ for the system can also be written in terms of eigen-modes as follows:
 \bea
 \{p\} = \sum_{i = 0}^{2 N_{\rm cells}-1} \beta_i \{\phi_i\},
 \eea
As can be seen from Eq.~\ref{eq:alphadot}, for the case of $\xi \approx \mu \lambda_1$, the velocity component corresponding to $\phi_0$ increases, whereas the components corresponding to other modes essentially decay to zero---in the very least they do not grow as fast as $\alpha_0$.  It may be noted that, since, the polarization vector for each cell has unit magnitude, the consolidated column vector will additionally satisfy
\bea
\label{eq:beta}
\langle p \rangle\{ p \} = \sum_{i = 0}^{2 N_{\rm cells}-1}\beta_i^2 = N_{\rm cells}.
\eea
Hence, the fact that $\dot{\alpha}_0$ is the dominant mode will ensure that $\beta_0$ will increase to some bounded steady state value that would depend on the polarization orientation parameter $\xi$, since, as per our polarization rule the polarization of a cell tends to align with its velocity. It may, however, be noted that since the polarization of each cell is a unit vector, in addition to a dominant $\beta_0$, some other $\beta_i$ components would also remain non-zero. This means that, as per Eq.~\ref{eq:alpha2}, some energy will keep getting pumped in a few other modes $i$. Nonetheless, $\dot{\alpha}_0$ will be the only component that would increase steadily as per Eq.~\ref{eq:alphadot}---other components $\dot{\alpha}_i$ would decay.

We now examine two extreme limits of polarization orientation constant $\xi$. When $\xi$ is very small($\xi << 1$), the response of $\hat{\mathbf p}$ to velocity ${\mathbf v}$ is slow. As a result, the
polarizations of the cells would lag behind in their bid for orienting with the velocity (see Fig.~2(a), (b), (d) of the main paper), resulting in a smaller steady state value for $\beta_0$. However, as described above, even a small component $\beta_0$ of the polarization field would be sufficient to sustain  steady $\dot{\alpha_0}$, and hence rotation---the other modes $\dot{\alpha}_j$  would not be sustained despite having non-zero $\beta_j$ values in some of the modes. The angular velocity of the tissue ($\omega \approx v_0 \beta_0$) would be, of course, small as is seen in Fig.~2(c). In the other extreme limit when $\xi \gg 1$, the rate at which energy is pumped in the modes is faster than the rate at which it is typically dissipated for the $j^{\rm th}$ mode~($\sim \mu \lambda_j$). In this case, we can see (Video S3), a perfect rotation of the tissue about the center is not obtained---the centre of rotation keeps changing constantly, confirming that $\{\phi_0\}$ is not the only mode that is invoked. Indeed, as can be seen from Video S3, a few radial modes are also excited, and are reminiscent of such movements observed in Ref.~\cite{deforet2014}. Our calculation predicts that, if the cells are highly cohesive, in which case their polarization can evolve faster~\cite{vitorino2008modular}~($\xi \gg 1$), we are expected to see these non-rotational, radial, modes.  Similarly, since the stiffness of the long wavelength modes ($\lambda_j$) is inversely proportional to the system size (e.g., in $1-D$, $\lambda_j \sim j^2/L$), we can see such modes for larger system even if $\xi \approx 1$. In fact, such movements are also reported in the experiments of Deforet et al \cite{deforet2014} and the simulations of Ref.~\cite{li2014coherent} for larger system sizes---the authors attribute these modes to the lack of the strength of persistent force ($v_0/\mu$ in our case). Our calculation, however, gives a clearer understanding for the origin of these movements.

The previous argument hinged on the tissue having a well defined stiffness matrix $[K]$, which in turn depends on having a system of cells with fixed connectivity. However, in our model we allow for the cells to change their neighbors and release internal stress. If that happens, the stiffness matrix of the tissue gets modified to $[K']$, depending on the rate at which the cells change their neighbors. As a result, the eigenvectors of the system now get modified to $\{\phi_i'\}$. The only eigenvector that is, however, most certainly common to the two systems is the one corresponding to the rigid body rotation $\{\phi_0\} = \{\phi_0'\}$.  Consequently, though there will be perturbations to $\beta_i$ in the form of new $\beta_i'$, the steady pumping of energy to the rotational mode will continue. The system is, hence,  expected to achieve rotational coherence even if the cells (nodes) are allowed to change neighbors. Though this argument is not  rigorous, it is consistent with the results of our simulations, and indeed seems plausible.

There are a couple of things that we did not account in the above derivation: (i) pre-stress
in the system due to crowding, and (ii) finite rotation effects. The pre-stress effect is too complex to
be accounted for in this simple derivation---we leave it for future work. The effect of finite rotation seems to be a
secondary effect. We see (Video S1) that when the cell sheet, by and large, behave elastically the coherent rotation is
initiated before any finite rotation actually happens in the system. As a result, we think that the finite rotation effect
is secondary, and if required can be incorporated by moving to a co-rotational frame of reference (as is done in section
deriving the analytical solution for elastic solids in main text)---it should not affect essential mechanics of the problem.

\subsection*{Text S3 --
Supplemental text showing the details of exact steady state solution when the tissue is
a viscous fluid.}

Since, in our model, the cells are allowed to change neighbours and relax the internal stress, then depending on the internal strain/stress, the tissue can indeed behave more like a fluid than like an elastic solid as described in the main paper. We hence, present a simple semi-analytical case of a Newtonian fluid to demonstrate coherent rotation for a fluidised tissue, and resort to the simulation results to make any contact with experiments. A more realistic, description of the tissue as a complex fluid is beyond the scope of this paper, due to the difficulty in both, using an appropriate rheological model, as well as in obtaining analytical solutions.

In this case, we seek to obtain a radially symmetric solution such that both the polarisation $\phat$ and velocities are aligned along the tangential direction (i.e., $v_r = 0$). To do so,  we write a very simple form for the equation of equilibrium as is given below. The equation for polarisation evolution remains the same as before (Eq. S1), and if we can find a such a solution, then we have found one possible steady state solution.

The constitutive equation for the epithelial sheet that is modelled as a viscous fluid is written as
\bea
\sigma = \eta \left (  \nabla \vel  + (\nabla \vel)^T -  (\nabla \cdot  \vel )\mathds{1} \right) + \eta_v (\nabla \cdot  \vel )\mathds{1},
\eea
%
%
where $\eta$ is the $2-D$ shear viscosity, and $\eta_v = 3 \eta$ in $2-$D~\cite{vkumar2014}. The equation of equilibrium in radial direction will be trivially reduced to zero for the solution that we are looking for. In the $\theta$ direction, the equation for tangential velocity $v_\theta$ will become,
\bea
2\eta \left ( {1 \over r}{\partial \over \partial r}\left (  r {\partial v_\theta \over \partial r}\right ) - {v_{\theta} \over r^2} \right)  + {1\over \mu_s} (v_0 - v_\theta) = 0,
\eea
with the boundary conditions
\bea
v_\theta(0) = 0,\mbox{ (symmetry) }\;\;\;\;{\partial \over \partial r}\left ( {v_\theta \over r} \right )_R = 0 \mbox{ (shear traction at boundary)}.
\eea
This equation seems to have a complicated closed-form solution in terms of Bessel and Hypergeometric functions. However, we can solve this problem numerically. To do that we will first non-dimensionalize the equation: all velocities are expressed in terms of $v_0$ and all lengths in terms of $R$. The equation then simplifies to
\bea
\label{fldeqn}
{1 \over r} {\partial \over \partial r}\left (  r {\partial v_\theta \over \partial r}\right ) - {v_{\theta} \over r^2}   + \alpha (1 - v_\theta) = 0,\;\;\;\;\;\left(\alpha = {R^2 \over 2 \mu_s\eta }\right).
\eea
The quantity $\sqrt{\mu_s \eta} = R_h$ is the hydrodynamic length~\cite{vkumar2014}, and the ratio $\alpha$ is the relative size of confinement disc ($R$) with respect to the hydrodynamic length $R_h$.   The non-dimensional Eq.~\ref{fldeqn} equation can be easily solved numerically. For low values of $\alpha$ the solution seems to be very similar to a rigid body rotation. When $\alpha$ becomes larger, the velocity initially increases with $r$ and then saturates to $v_0$. One such plot for $v_\theta$ with respect to $r$ is shown in Fig.~S2.

\subsection*{Text S4 --
Supplemental text showing that passive confinement is also capable of inducing coherent rotation.}
In order to test whether cells can coherently rotate under passive confinements
instead of geometrical confinements, simulations were performed with three
initial positions of the `active' cells: in the center of the circular pattern
(Figs.~S4 (a)--(c), Video S4--S5), in the periphery of the circular pattern
(Figs.~S4 (d)--(f), Video S6--S7) and in an annular geometry
(Fig.~S3, Video S8--S11). Consistent with our hypothesis, our simulations with
varying mobility ratios of active and passive cells suggests that differences in
physical properties of cells (i.e., motility and mobility) can indeed induce
coherent angular motion (Fig.~S3, Video S8 - S11). When the
frictional properties of active and inactive cells were comparable
(i.e., $\mu_{\rm active}/\mu_{\rm inactive}) = 1$, active cells were found
to intercalate into the tissue and not exhibit any coherent motion. However,
a 100-fold decrease in mobility of passive cell induced segregation of active
cells due to inability of active cells to move the passive cells, and led to
onset of coherent motion. This was even more clear for $\mu_{\rm active}/\mu_{\rm inactive} = 1000$ where
the majority of the active cells remained stuck in their initial positions, i.e., either centrally (Fig.~S4(c)),
or peripherally (Fig.~S4(f)) or along the annulus (Fig.~S3(d)), and exhibited coherent angular motion.
Together, these results indicate that passive confinement effected by differences in cell mechanical properties is sufficient to induce
coherent rotation.

\subsection*{Text S5 --
Supplemental text showing a detailed analysis to understand mechanisms that govern fluidisation of the tissue during coherent angular movement.}
We had addressed various parameters and mechanisms that can lead to fluidisation of the tissue in Section ``Cell crowding leads to fluidisation of tissue.'' Here, we present a detailed analysis of those mechanisms. Note that in the discussion below we interchangeably use distortion and shear strain; distortion implies change in shape, and shape deformation is the hallmark of shear strain~\cite{sadd2009}.

\subsection*{Confinement contributes to the distortion of the tissue}
In the absence of any compression, the approximate size of a single undeformed cell in a tissue, assuming the cell shape to be approximately circular,  will be $\pi (a_0/2)^2$. As a result, the critical number of cells that can be approximately packed in a circle of radius $R$ is $N_c \approx R^2/(a_0/2)^2$. Using the canonical values for our model, $R = 5$ and $a_0 = 1$, this number $N_c \approx 100$. Hence, in a loose sense, a tissue under this confinement with more than $100$ cells is expected to have deformation. If the tissue were elastic and homogeneous, one would expect the confinement leading simply to uniform and isotropic compression~\cite{sadd2009}. However, due to its discrete nature, the tissue is susceptible to having shear deformations in addition. This is because,  the circular confinement does not permit all the cells to have a preferred co-ordination number $z = 6$~\cite{williams2014}, thus leading to distortion of the tissue, since the triangles formed by cell-cell connections (springs) cannot be all equilateral. The presence of motility forces further distorts the tissue by contributing additional shear strains. In the steady state of coherent rotation, we hence, expect the distortion of cell-connection triangles to depend on cell density. We will see below that, these intuitive observations are indeed consistent with the findings of our simulations.

As described in the main paper, the connectivity of cells is obtained by Delaunay triangulation~\cite{mirams2013chaste}.  We can plot histograms of the qualities of cell-connection triangles spanning the tissue in Fig. S7. The quality of a triangle is defined as~\cite{matlab2012}:
\bea
Q = \frac{4\sqrt{3} A}{h_1^2 + h_2^2 + h_3^2},
\eea
where, $h_1, h_2$ and $h_3$ are the side lengths of any given triangle, and $A$ is its area. The quality factor $Q = 1$ for an equilateral triangle (no distortion), and reduces with increasing distortion.  It can be clearly seen from Fig.~S7 (a) and (b) that the number of relatively distorted triangles $Q < 0.9$ (chosen arbitrarily) is larger for the case $N = 170$ when compared with the less crowded $N = 140$. The steady state shear strain is, hence, observed to be comparatively larger for the more crowded tissue.

 Delaunay triangulation  updates the connectivity of cells  if the updated triangulation has fewer ``skinny," i.e, distorted triangles~\cite{delaunay}. Hence, in a statistical sense, a more distorted tissue has greater susceptibility for changing its connectivity, thus resulting in its fluidization.

Since Delaunay triangulations are dual to  Voronoi tessellations~\cite{delaunay} (Fig.~1(b) of the main paper), cell distortions can be equivalently but better explained by looking at the statistics of Voronoi (cell) edge length of the tissue in Fig.~S8 (a) and (b) for $N = 140$ and $N = 170$.  Cells connected to each other share an edge. When the cells update their connectivity via Delaunay triangulation, this edge grows in the topologically orthogonal direction while transitioning through a point, or a four-way vertex (see Fig.~S9). Hence, in a statistical sense, a tissue with a substantial number of small edges is  more susceptible to T1 transitions (or neighbour change). Since, it can be clearly seen from Fig.~S8 that, in its steady rotating state, the number of very small Voronoi edges $e < 0.2$ (chosen arbitrarily)  for $N = 170$ case, is quite large when compared with the case $N = 140$, we have another cue as to why the tissue is more prone to fluidisation for $N = 170$, when compared with $N = 140$.

\subsection*{Increase in cell-cell connection (spring) stiffness reduces distortion and makes the tissue resistant to fluidisation}
Upon increasing the stiffness of springs, though the initial shear pre-strain is not expected to change much due to the kinematic confinement provided by the circular boundary, the additional distortion caused by the motile forces would clearly be decreased. The decrease in overall distortion of the tissue for $N = 170$ can be clearly seen from Figs.~S7(c)~and~S8(c). Due to the relative lower distortion, as compared to a softer tissue, the tissue is  more resistant to neighbour changes and hence fluidisation.

\subsection*{Neighbour changes via Delaunay triangulation happen in local patches in a time-sequential manner and releases local shear strain in the tissue}
For any configuration of cell centers, Delaunay triangulation, in essence, provides the cells with connectivity that produces the least number of distorted cell triangles. As demonstrated in Fig.~S9 (also see SI Video~S22), for $N = 170$, the neighbour exchange happens at local ``hot spots," in a frequent, and time-sequential manner.  A local T1 transition event release the local shear strains in the tissue, as and when the strain builds up, and a neighbour change can reduce the same.
 As also described in the SI section below,  T1 transition happens in our model only if re-triangulation results in a tissue with lesser distortion.  If we prevent the T1 transition from happening, by ``locking" the connectivity of cells, i.e., by not updating the connectivity of the cells, then distortion starts building up at local patches (SI Fig.~S10(b) for $N  = 170$) and steadily increases. Updating the tissue connectivity by Delaunay triangulation, which leads to fluidisation of the tissue, prevents build-up of shear strain and corresponding stress in the tissue. Thus updating cell-cell connectivity via Delaunay triangulation is a mechanically very relevant way to release shear strains in a tissue. Note that ``locking" the connectivity does not lead to building up of stress for $N = 140$ (SI Fig.~S10(a) and SI Video~S21) because any modification in connectivity would only increase distortion in the tissue.

\subsection*{ Skewness of cell-cell connection length distribution is a key quantity that decides the susceptibility of the tissue to fluidisation}
The description given in the previous section clearly demonstrates the correlation between the shear distortion caused in the tissue by confinement with fluidisation upon crowding in the tissue. However, this does not explain why fluidisation is seen for $N = 170$ and not for $N = 140$, though confinement related distortion is present in both cases (see Fig.~S7). Below we clearly describe that though the presence of shear strain is necessary for T1 transitions in the tissue,  in itself it is not sufficient. For T1 transition to happen it is important  that the local distortion of cells reduces upon T1 transition.  Indeed, for $N = 140$, there are triangles with distortion comparable for the case $N = 170$ (see Fig.~S7). What then is the possible reason for the stability of $N = 140$, with respect to T1 transitions, despite having distortion in triangles (albeit fewer numbers) that is comparable to $N = 170$? The resolution to this question, again, lies with the degree of confinement. For $N  = 140$, since the density of cells is relatively lower, in addition to short edges we expect to find a chunk of long connections with length $\approx a_0 = 1$ at steady rotation state. The presence of such relatively long edges amongst short edges would lead to triangles of the type shown in Fig.~S11(b). Though these triangles indeed have shear distortion, connectivity update will lead to triangles of even higher distortion. When the number of cells is increased, the proportion of longer edges decreases, resulting in triangles shown in Fig.~S11(b) that are not as robust to triangulation update.

To quantify, the proportion of long connections to short connections in the tissue we obtain the quantified the skewness in the distribution of edges for $N = 140, 150, 160, 170$ given by Pearson's second coefficient $ Sk_2$ given by~\cite{sk2011}:
\bea
Sk_2 = 3 \frac{(\rm{mean} - \rm{median})}{\rm{standard\;deviation}}.
\eea
As shown in Fig.~S11(c),  $Sk_2$ increases with cell crowding, i.e., greater the skewness in the connection distribution, greater the amount of fluidisation of the tissue. This correlation is borne out consistently in other situations where shear can happen in the absence of crowding.
\begin{enumerate}
\item Consider the case, when $R = 10$ and $N = 560$, this corresponds to cell density equivalent to $R = 5$ and $N = 140$. So, contrary to our exception, we indeed observe fluidisation of the tissue during steady rotation. This is very likely because the maximum shear in the tissue is proportional to $R$ (see Eq.~14 and Fig.~5(a) of the main paper), as a result of which, the distortion due to motile forces can push the tissue over triangulation stability. The same behaviour is observed for $R = 15$ and $N = 1080$ (Fig.~5(a)). The index $Sk_2$ is around $0.4$ for both cases, which is greater than the value of $Sk_2 = 0.3$ for $R = 5$ and $N = 140$ (no fluidisation).
\item If we increase the radius of confinement to $R = 5.8$ such that the cell density for $N = 170$ is similar to the case $R = 5$ and $N = 140$, we indeed observe that there is no fluidisation in the tissue. The value of $Sk_2$ reduces from $0.54$ ($N = 170, R = 5$) to $0.16$.
\item If the stiffness of the cell connections is increased to $k_c = k_t = 100$, the  fluidisation of the tissue is prevented and the  value of $Sk_2$ reduces drastically from $0.54$ to $0.04$. On the other hand, if $k_t$ is reduced to $1$, then even for $N = 140$ and $R = 5$ case, tissue undergoes fluidisation, and the value of $Sk_2$ increases from 0.3 for no fluidisation to $0.4$.
\end{enumerate}
To summarise the findings, we find that the skewness ratio of connection lengths is intricately linked with the susceptibility of tissue to fluidisation.

\subsection*{Text S6 --
Supplemental text showing the details of Delaunay Triangulation and T1 transitions in our model.}
In our model, the cell connectivity is obtained via Delaunay triangulation from cell centre positions at every time step \cite{li2014coherent, pathmanathan2009computational}. The connectivity of the cells is only modified when there is local shear distortion of the tissue, which can be typically relieved by a systematic exchange of neighbour pairs (Figs.~S9 and S10). These modification of neighbours that can be interpreted as the so called T1 transitions, the specialised terminology for neighbour exchange in the context of foams and epithelia~\cite{fletcher2014vertex}. T1 transitions that are dependent on shear deformation of cells, have been also observed/modelled in a very recent work by Etournay et al. \cite{etournay2015}, thus providing biological pointer for distortion driven neighbor changes in a tissue. It may be noted that, in the case of Vertex Models (VM), that have been extensively used for modelling epithelial morphology during development, the T1 transitions typically happen when the edge-length shared by two-cells reduces beyond an arbitrarily chosen critical threshold, upon which the connectivity is updated~\cite{fletcher2014vertex}. The existence of a small edge in a cell, naturally implies the presence of shear strain, and is quite in line with distortion criteria used by Delaunay triangulation (and the corresponding Voronoi tessellation). However, Delaunay triangulation has one advantage of updating the cell connectivity naturally without resorting to potentially ad-hoc assumption used by VM \cite{jennings2014new, wyatt2015emergence}.

 \begin{figure}[H]
 \includegraphics[width=11.4 cm]{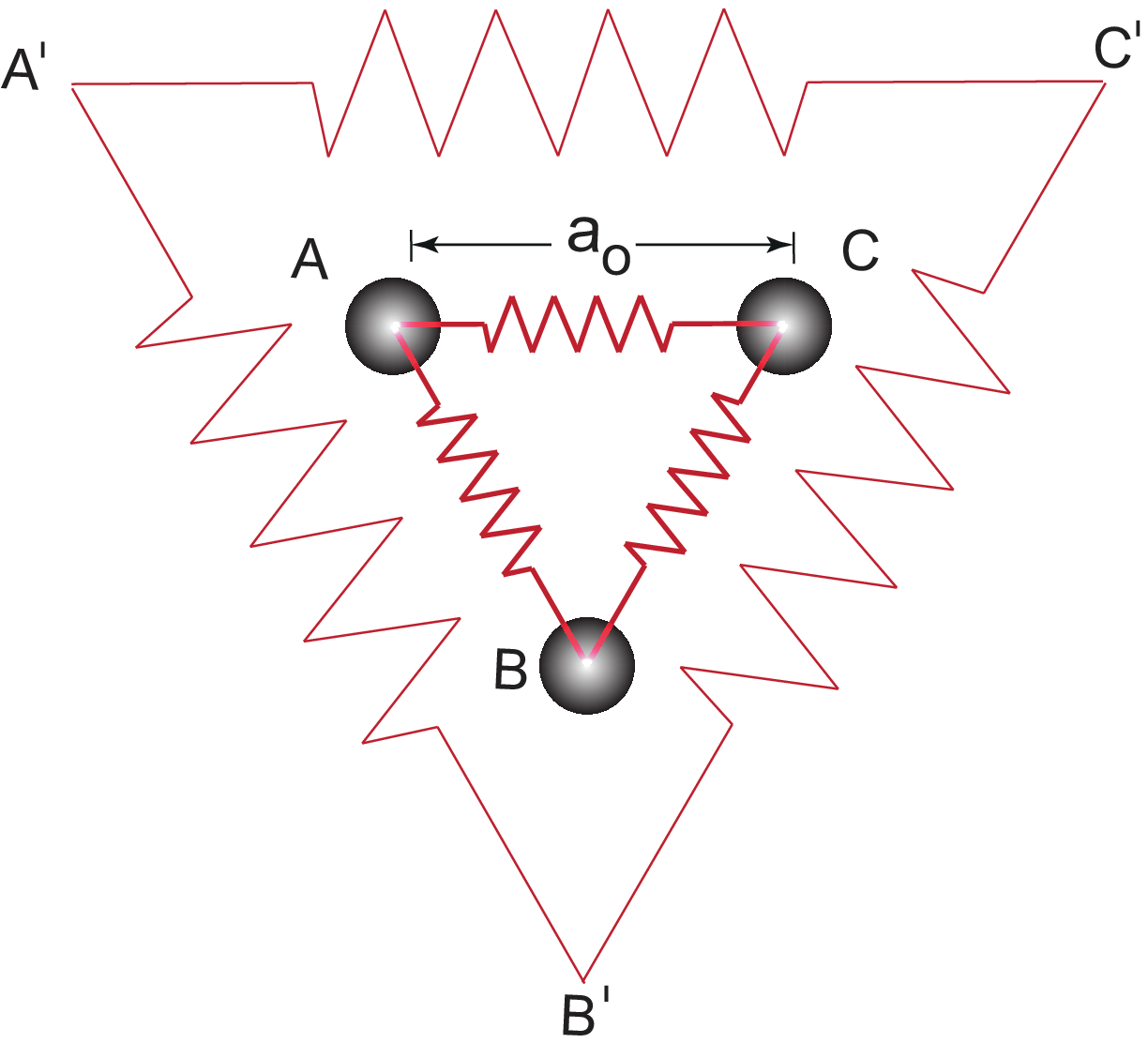}
 \begin{centering}
 \caption{\label{supplifig:1} {\bf Triangular network of cells}
 }
 \end{centering}
 \end{figure}
\begin{figure}[H]
\includegraphics[width=11.4 cm]{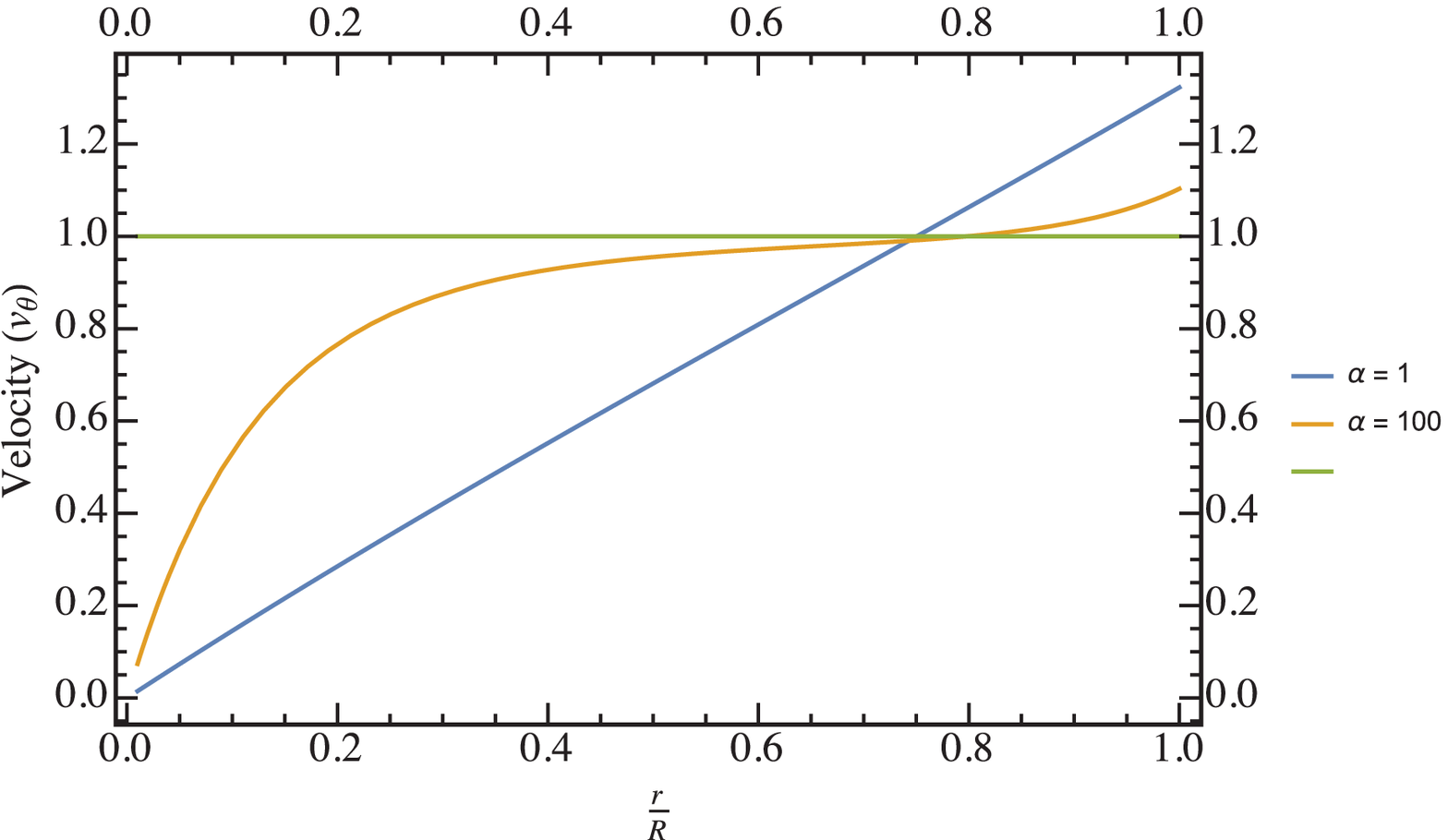}
\begin{centering}
\caption{
\label{visco}{\bf Numerical solution for tangential velocity
for $\alpha = 1 \mbox{ and } 100$.} When $\alpha = 1$, i.e., $R_h  \approx R$, the tissue rotates almost rigidly. On the other hand, when $\alpha = 100$, i.e., when $R \gg R_h$, then the velocity increases with $r$ and then saturates to value close to $v_0$.
}
\end{centering}
\end{figure}
 \begin{figure}[H]
 \includegraphics[width=11.4 cm]{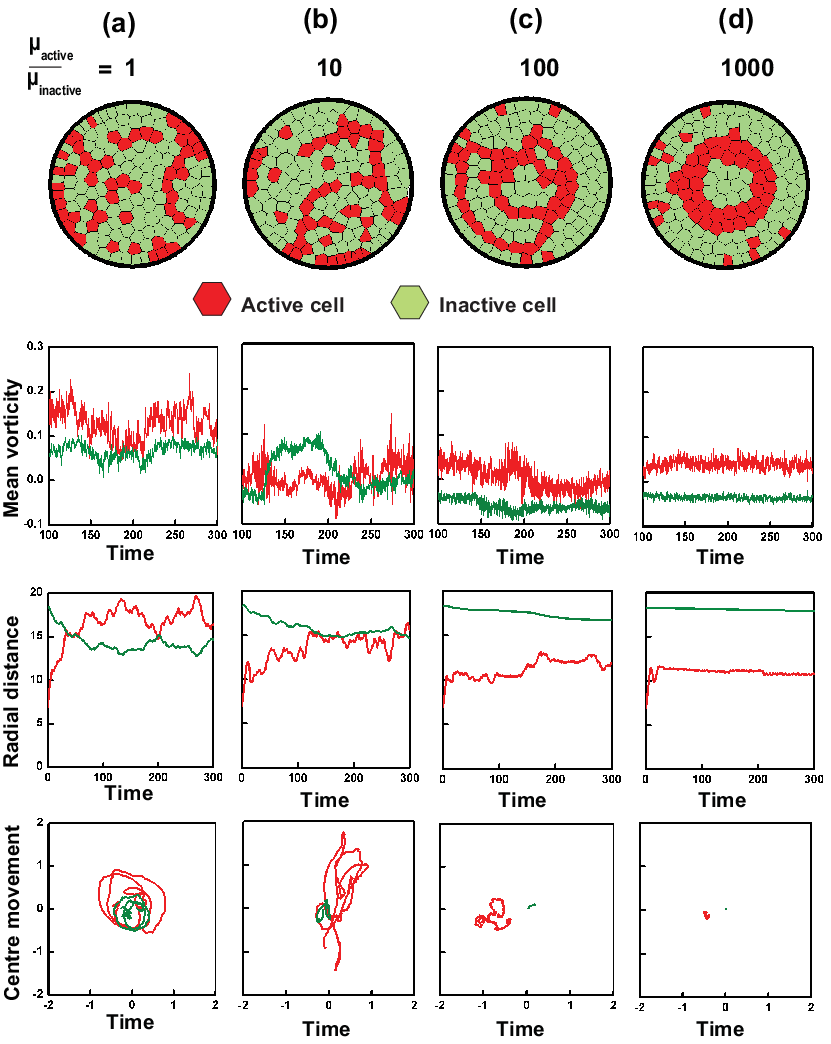}
 \begin{centering}
 \caption{\label{supplifig:2} {\bf Passive confinement inducing coherent rotation.} {(a) A tissue comprising of heterogeneous populations of cells with active (red) and passive cells (green) undergo coherent motion depending upon the relative properties of two populations. Active cells, which are initially embedded in the form of an annular ring in passive tissue try to intercalate into the tissue, if their mobility ratios are comparable. With decrease in passive cell mobility (increase in friction), active cells experience difficulty in penetrating out from their initial position which results in a robust coherent motion of active cells.
(a) - (d) show heterogeneous tissues with different mobility ratios ranging from 1 to 1000. In order to estimate the rotational motion of cells in the heterogenous tissue, we choose mean vorticity as the quantifying parameter as in earlier cases. For lower mobility ratios, active cells try to  drag passive cells also along with them and try to build up a total rotation of the system. As the mobility ratio increases, their ability to drag passive cells decreases and for highest mobility ratio, active cells exhibit persistent rotation with steady value of mean vorticity.
The spreading characteristic of cells from their equilibrium position is measured using mean radial distance and center movement. }
}
 \end{centering}
 \end{figure}

 \begin{figure}[H]
 \includegraphics[width=11.4 cm]{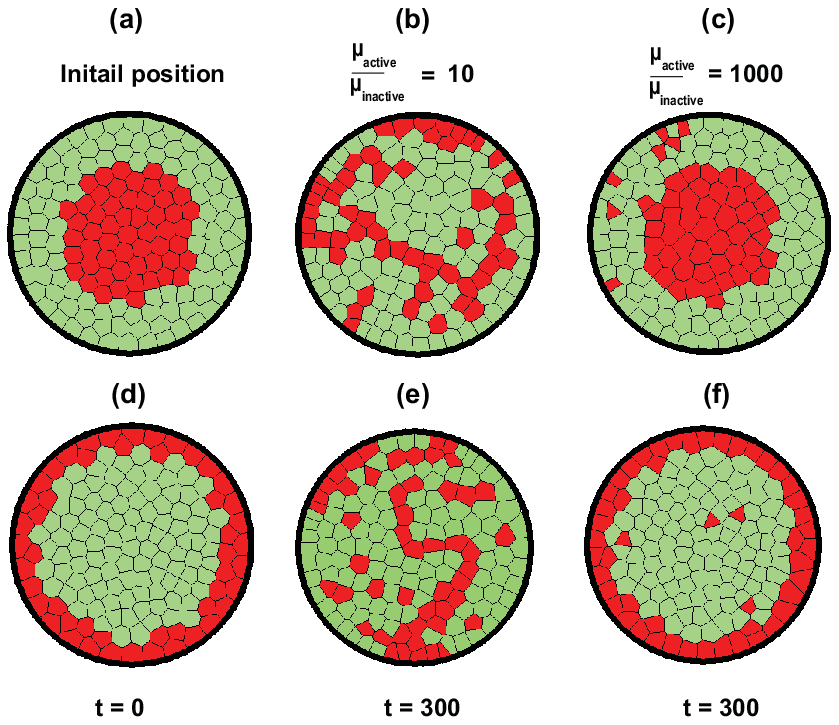}
 \begin{centering}
 \caption{\label{supplifig:3} {\bf Different positions of active cells in passive tissue.} (a) - (c)
show cells placed at the center and (d) - (f) show the cells
placed at the periphery of passive tissue. Simulations with two different mobility
ratios (10 and 1000) show that system behavior is the same as that obtained for
annular positioning of active cells. As in previous case, here also we see that for lower
mobility ratio, active cells try to intercalate into the passive tissue and drag the neighboring
passive cells along them. Similarly, for higher mobility ratio, active cells
do not penetrate much outside their initial position, instead they exhibit
coherent rotation.
}
 \end{centering}
 \end{figure}
\begin{figure}[H]
\includegraphics[width=6 cm]{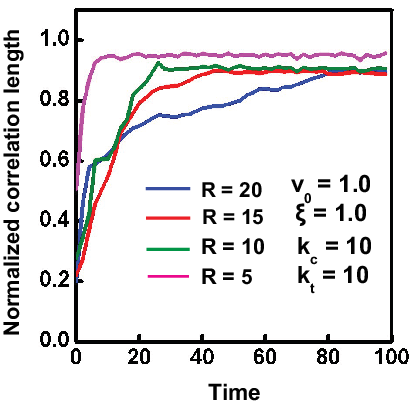}
\begin{centering}
\caption{
\label{NormCorrelation}{{\bf Correlation length normalized with system size is plotted for increasing time}. As the size of the tissue increases, the time taken to reach coherence also increases.}
}
\end{centering}
\end{figure}

\begin{figure}[H]
\includegraphics[width=11.4 cm]{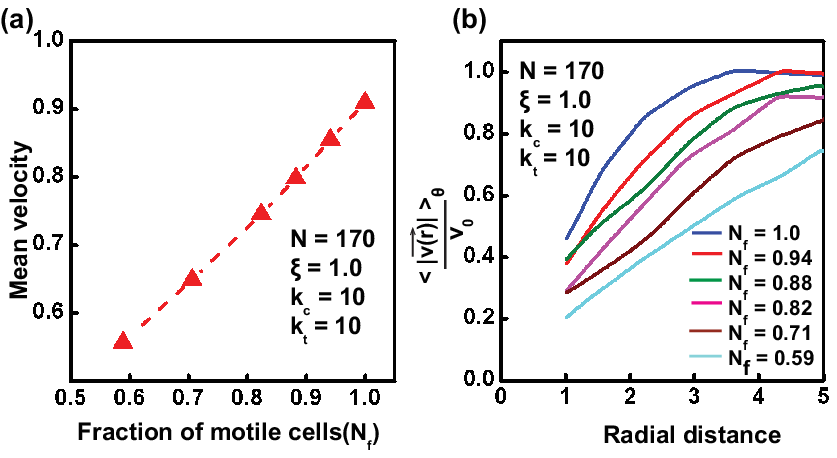}
\begin{centering}
\caption{
\label{motilefrac}{{\bf Velocity profile when only a fraction of cells in tissue are motile.} (a) With decrease in fraction of motile cells, the mean velocity of system reduces. (b) Velocity profile for varying fraction of motile cell shows that as the number of motile cells increases, system behavior will change from solid-like to fluid-like.}
}
\end{centering}
\end{figure}
\begin{figure}[H]
  \includegraphics[width=11.4 cm]{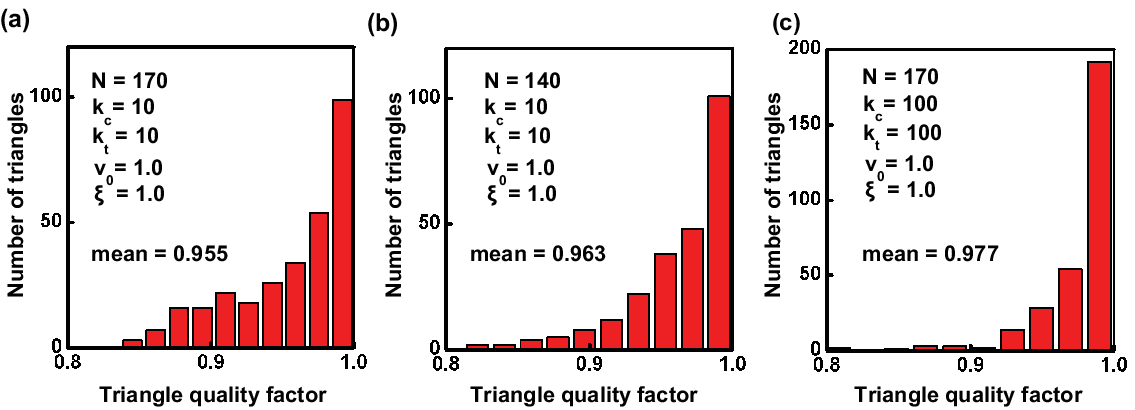}
  \centering
  \caption{{\bf Quality factor for cell triangles of a coherently rotating confined tissue for different physical conditions}.  All other parameters remaining the same, it can be clearly seen that, (a-b) the proportion of relatively more distorted triangles is larger for the denser, fluidised, tissue  ($N = 170$) as compared to a less dense tissue  ($N = 140$). (c) When the stiffness of cell-cell connections (springs) is increased, the number of distorted cell triangles in the tissue  with $N = 170$ is significantly reduced. As described in the main paper, the tissue now coherently rotates like a solid. }
  \label{triQ}
\end{figure}
\begin{figure}[H]
   \includegraphics[width=11.4 cm]{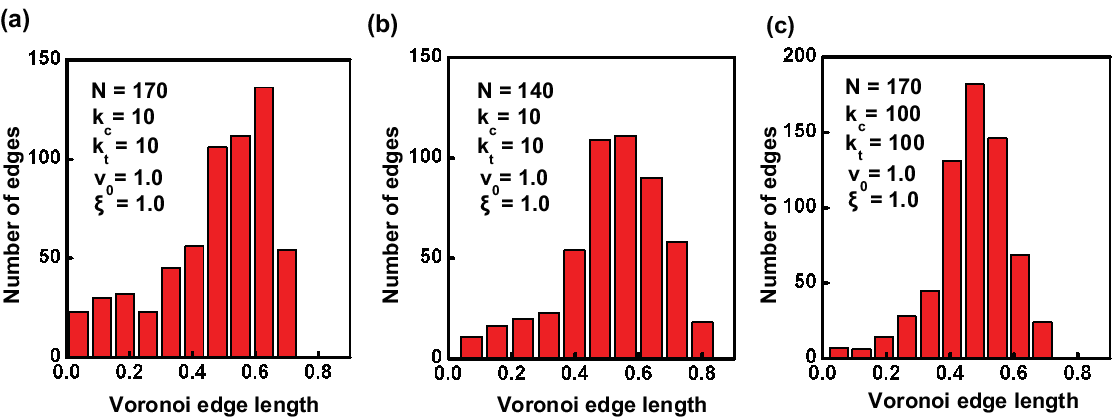}
  \centering
  \caption{{\bf Voronoi (cell) edge length distribution for different tissue parameters}. All other parameters remaining the same, it can be clearly seen that, (a-b) the proportion of relatively small edge lengths is larger for the denser, fluidised, tissue  ($N = 170$), as compared to a less dense tissue ($N = 140$). (c) When the stiffness of cell-cell connections (springs) is increased, the number of small edges in the tissue  with $N = 170$ is significantly reduced. As described in the main paper, the tissue now coherently rotates like a solid.}
  \label{vorEdge}
\end{figure}
\begin{figure}[H]
   \includegraphics[width=11.4 cm]{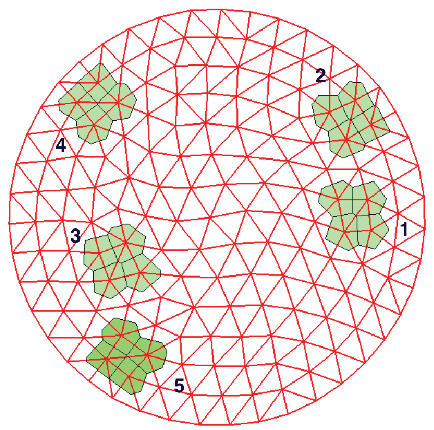}
  \centering
  \caption{ During coherent rotation of a tissue with higher cell density ($N = 170$), the neighbor changes happen in local patches with relatively higher distortion that can be reduced upon local connectivity update via Delaunay triangulation. The corresponding Voronoi cells at these places are also shown. It can be seen that such patches keep appearing locally at different places in a time-sequential manner (also see the related SI Video~S20). }
  \label{T1}
\end{figure}
\begin{figure}[H]
   \includegraphics[width=11.4 cm]{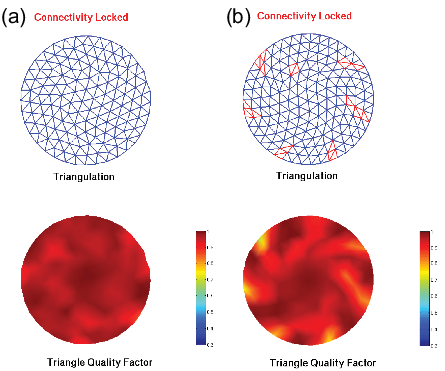}
  \centering
  \caption{(a) ``Locking" the node connectivity for less dense tissue ($N = 140$)  does not modify the mechanical state of the coherently rotating tissue (see SI Video~S21). (b) On the other hand, ``locking" the connectivity of denser tissue ($N = 170$) leads to build up of distortion (red triangles) in the tissue. It can be seen from SI Video~S22 that, upon ``releasing" the connectivity lock, many cells undergo neighbor changes to relieve their distortion (shear). }
  \label{triQ170}
\end{figure}
\begin{figure}[H]
  \includegraphics[height=4cm]{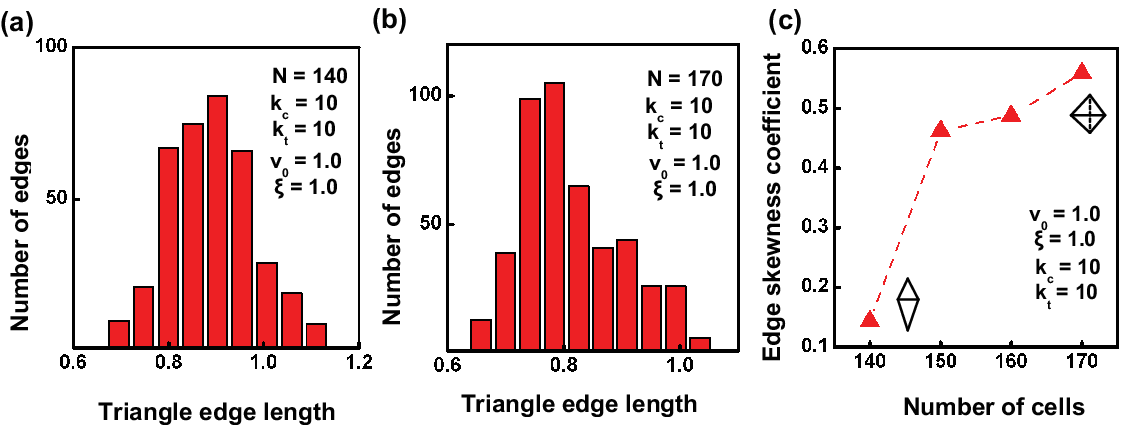}
  \centering
  \caption{(a-b) It can be seen that the distribution of cell-cell connection (spring) lengths, which is a measure of distance between cell centers, becomes more skewed towards lower spring lengths, as the number density of cells increases. (c) The skewness as a function of cell numbers is quantified using Pearson's second skewness coefficient $Sk_2$. When the connection length distribution is less skewed, it implies that the number of large and small springs are comparable to each other. This may indeed lead to the presence of some distorted triangles edges as shown, but a connectivity update would lead to greater increase in distortion and hence not performed. The reverse is true when the distribution is more skewed.}
  \label{skew}
\end{figure}

\renewcommand{\figurename}{Video}
\setcounter{figure}{0}
\renewcommand{\thefigure}{S\arabic{figure}}
\begin{figure}[H]
\caption{\bf Emergence of coherent angular rotation of cells confined in a circular geometry.} The parameters for the simulations are
  $N= 140$, $k_c = k_t = 10$, $\xi =1$, $v_0=1$.
\end{figure}
\begin{figure}[H]
\caption{\bf Higher number density of cells leads to the fluidization of tissue, indicated by the huge shear
  appearing in the system.} The parameters for the simulations are
  $N= 170$, $k_c = k_t = 10$, $\xi =1$, $v_0=1$.
\end{figure}
\begin{figure}[H]
\caption{\bf For higher values of $\xi$, center of rotation keeps on changing.} The parameters for the simulations are
  $N= 140$, $k_c = k_t = 10$, $\xi =10$, $v_0=1$.
\end{figure}
  \begin{figure}[H]
\caption{\bf Active cells initially placed in circular
  pattern in the center of a passive tissue.} The parameters for the simulations are
  $N= 154$, $k_c =5$, $k_t = 1$, $\xi =1$, $v_0$(active cells)=1, mobility ratio = 10.
\end{figure}
\begin{figure}[H]
\caption{\bf Active cells initially placed in circular
  pattern in the center of a passive tissue.} The parameters for the simulations are
  $N= 154$, $k_c =5$, $k_t = 1$, $\xi =1$, $v_0$(active cells)=1, mobility ratio = 1000.
\end{figure}
\begin{figure}[H]
\caption{\bf Active cells placed on the periphery of a passive
  tissue in a circular pattern.} The parameters for the simulations are
  $N= 154$, $k_c =5$, $k_t = 1$, $\xi =1$, $v_0$(active cells)=1, mobility ratio = 10.
\end{figure}
  \begin{figure}[H]
\caption{\bf Active cells placed on the periphery of a passive
tissue in a circular pattern.} The parameters for the simulations are
  $N= 154$, $k_c =5$, $k_t = 1$, $\xi =1$, $v_0$(active cells)=1, mobility ratio = 1000.
\end{figure}
  \begin{figure}[H]
\caption{\bf Passive confinement inducing coherent rotation
in system with active cells placed inside a passive
tissue in an annular pattern.} The parameters for the simulations are
  $N= 154$, $k_c =5$, $k_t = 1$, $\xi =1$, $v_0$(active cells)=1, mobility ratio = 1.
\end{figure}
  \begin{figure}[H]
\caption{\bf Passive confinement inducing coherent rotation
in system with active cells placed inside a passive
tissue in an annular pattern.} The parameters for the simulations are
  $N= 154$, $k_c =5$, $k_t = 1$, $\xi =1$, $v_0$(active cells)=1, mobility ratio = 10.
\end{figure}
  \begin{figure}[H]
\caption{\bf Passive confinement inducing coherent rotation
in system with active cells placed inside a passive
tissue in an annular pattern.} The parameters for the simulations are
  $N= 154$, $k_c =5$, $k_t = 1$, $\xi =1$, $v_0$(active cells)=1, mobility ratio = 100.
\end{figure}
  \begin{figure}[H]
\caption{\bf Passive confinement inducing coherent rotation
in system with active cells placed inside a passive
tissue in an annular pattern.} The parameters for the simulations are
  $N= 154$, $k_c =5$, $k_t = 1$, $\xi =1$, $v_0$(active cells)=1, mobility ratio = 1000.
\end{figure}

 \begin{figure}[H]
\caption{\bf Emergence of coherent motion in annular geometry.} The parameters for the simulations are
  $N= 100$, $k_c =10$, $k_t = 10$, $\xi =1$, $v_0$ = 1.
\end{figure}
 \begin{figure}[H]
\caption{\bf The pattern of coherent motion is determined by the stiffness of cell -cell connection.} The parameters for the simulations are
  $N= 100$, $k_c =1$, $k_t = 1$, $\xi =1$, $v_0$ = 1.
\end{figure}
 \begin{figure}[H]
\caption{\bf Asynchronous rotation does not switch the direction of rotation.} The parameters for the simulations are
  $N= 40$, $k_c =5$, $k_t = 1$, $\xi =1$, $v_0$ = 1.
\end{figure}
 \begin{figure}[H]
\caption{\bf Synchronous rotation switches the direction of rotation.} The parameters for the simulations are
  $N= 40$, $k_c =5$, $k_t = 1$, $\xi =1$, $v_0$ = 1.
\end{figure}
 \begin{figure}[H]
\caption{\bf  Breakage of boundaries for softer system.} The parameters for the simulations are
  $N= 100$, $k_c =1$, $k_t = 1$, $\xi =1$, $v_0$ = 1.
\end{figure}
 \begin{figure}[H]
\caption{\bf Breakage of boundaries for medium stiff system.} The parameters for the simulations are
  $N= 100$, $k_c =10$, $k_t = 1$, $\xi =1$, $v_0$ = 1.
\end{figure}
 \begin{figure}[H]
\caption{\bf Breakage of boundaries for medium stiff system.} The parameters for the simulations are
  $N= 100$, $k_c =10$, $k_t = 1$, $\xi =1$, $v_0$ = 1.
\end{figure}
 \begin{figure}[H]
\caption{\bf Coherent rotation of cells in larger systems.} As the confinement radius R for the tissue increases, radial movements become prominent. The parameters for the simulations are
  $R = 15$, $N= 1260$, $k_c =10$, $k_t = 10$, $\xi =1$, $v_0$ = 1.
\end{figure}
 \begin{figure}[H]
\caption{\bf T1 transition happening in a denser system at local patches.} The parameters for the simulations are
  $N= 170$, $k_c =10$, $k_t = 10$, $\xi =1$, $v_0$ = 1.

\end{figure}
 \begin{figure}[H]
\caption{\bf “Locking” the node connectivity for less dense tissue does
not modify the mechanical state of the coherently rotating tissue.} The parameters for the simulations are $N= 140$, $k_c =10$, $k_t = 10$, $\xi =1$, $v_0$ = 1.
\end{figure}

 \begin{figure}[H]
\caption{\bf  “Locking” the connectivity of denser tissue (N = 170) leads to build
up of distortion (red triangles) in the tissue. Upon “releasing” the connectivity lock, many cells undergo neighbor changes to relieve
their distortion (shear).} The parameters for the simulations are $N= 170$, $k_c =10$, $k_t = 10$, $\xi =1$, $v_0$ = 1.
\end{figure}

\section*{Acknowledgments}
We acknowledge helpful discussions with H. Chat\'{e}, F. J\"{u}licher, D. Riveline, S. Ramaswamy and A. Sain. MMI acknowledges the hospitality from MPI-PKS, Dresden, where a part of this work was done. 
\nolinenumbers

%

\end{document}